\documentclass[aps,twocolumn,prd,showpacs,preprintnumbers,amsmath,amssymb,nofootinbib]{revtex4-1}

\pdfoutput=1

\usepackage{mathtools}
\usepackage{braket}				
\usepackage{graphicx}
\usepackage{xcolor}
\usepackage{dsfont}
\usepackage{units}						
\usepackage{csquotes}					
\usepackage{siunitx}
\usepackage{comment}
\usepackage{xspace}

\usepackage{hyperref}

\newcommand{\eg}{e.g.~}
\newcommand{\cf}{cf.~}
\newcommand{\ie}{i.e.~}

\newcommand{\hc}{\text{h.c.\xspace}}

\newcommand{\symhspace}[2]{\hspace{#1}#2\hspace{#1}}	

\newcommand{\fineq}[1]{\;{#1}}			
\newcommand{\sepeq}[1]{\; {#1} \quad}	
\newcommand{\myVec}[1]{\vec{#1}}			
\newcommand{\dd}{\mbox{d}}				
\newcommand{\transpose}{\intercal}			

\DeclarePairedDelimiter\chevron{\langle}{\rangle}
\DeclarePairedDelimiter\absVal{\vert}{\vert}

\DeclareMathOperator{\tr}{tr}				

\newcommand{\PlanckScale}{M_{\text{\tiny{Pl}}}}
\newcommand{\UVscale}{\Lambda_{\text{\tiny{UV}}}}
\newcommand{\GWscale}{\Lambda_{\text{\tiny{GW}}}}
\newcommand{\HiggsField}{H_{\text{\tiny{LHC}}}}
\newcommand{\mHiggs}{m_{\text{\tiny{Higgs}}}}
\newcommand{\mPGB}{m_{\text{\tiny{PGB}}}}

\newcommand{\BSM}{B_\text{\tiny{SM}}}
\newcommand{\BSMtilde}{\tilde{B}_\text{\tiny{SM}}}

\definecolor{darkblue}{rgb}{0.0,0.0,0.4}
\definecolor{darkgreen}{rgb}{0.0,0.4,0.0}
\hypersetup{
    colorlinks,
    linkcolor=darkblue,	
    citecolor=darkgreen,
    urlcolor=darkblue
}

\renewcommand{\eqref}[1]{Eq$. \,$(\ref{#1})}

\newcommand{\tinytext}[1]{\text{\tiny{#1}}}

\begin{document}


\title{Minimal conformal extensions of the Higgs sector}

\author{Alexander J. Helmboldt}
\author{Pascal Humbert}
\author{Manfred Lindner}  
\author{Juri Smirnov}

\affiliation{\vspace{0.5em}Max-Planck-Institut f\"ur Kernphysik, Saupfercheckweg 1, 69117 Heidelberg, Germany}

\pacs{}


\begin{abstract}
\noindent
In this work we find the minimal extension of the Standard Model's Higgs sector which can lead to a light Higgs boson via radiative symmetry breaking and is consistent with the phenomenological requirements for a low-energy realization of a conformal theory.
The model which turns out to be stable under renormalization group translations is an extension of the Standard Model by two scalar fields, one of which acquires a finite vacuum expectation value and therefore mixes into the physical Higgs.
We find that the minimal model predicts a sizable amount of mixing which makes it testable at a collider.
In addition to the physical Higgs, the theory's scalar spectrum contains one light and one heavy boson.
The heavy scalar's properties render it a potential dark matter candidate.
\end{abstract}

\maketitle


\section{Introduction}
\label{sec:intro}
\noindent
The gauge hierarchy problem continues to be one of the most pressing questions of modern theoretical physics. It is a naturalness problem which, at its core, asks the question why the electroweak scale can be light in spite of a high-energy embedding of the Standard Model (SM) into a more complex theory with other heavy scalar degrees of freedom. One approach to solve the hierarchy problem is the systematic cancellation of bosonic and fermionic loop contributions to the Higgs mass within supersymmetry. However, due to the fact that no supersymmertic particle has been observed yet, alternative approaches are appealing.

A radical way of addressing this problem is the assumption that the fundamental theory describing Nature does not have any scale. In such a conformal model, the symmetry can be realized non-linearly and explicit scales can appear. Early works that employ scale-invariant models to solve the hierarchy problem include \cite{Hempfling:1996ht,Meissner:2006zh,Meissner:2007xv,Foot:2007as,Foot:2007ay}.
In recent years those studies inspired a number of other works addressing different open questions beyond the SM \textendash\ like small neutrino masses, the nature of dark matter or baryogenesis \textendash\ in the context of scale-invariant theories; see for example \cite{Haba:2015qbz,Haba:2015nwl,Haba:2015lka,Karam:2015jta,Ahriche:2015loa,Humbert:2015yva,Kang:2015aqa,Carone:2015jra,Humbert:2015epa,
Lindner:2014oea,Hashino:2015nxa,Farzinnia:2015uma,Latosinski:2015pba,Okada:2014nea,Kang:2014cia,Benic:2014aga,Gorsky:2014una,
Khoze:2013oga,Hambye:2007vf,Carone:2013wla,Foot:2010et,Hill:2014mqa,Salvio:2014soa,Englert:2013gz,Alexander-Nunneley2010,Heikinheimo2014,Plascencia2015,Ghorbani2015b,Bertolini:2015boa}. A common feature of those works is the need for additional bosonic degrees of freedom, as in the SM alone the large top mass does not permit radiative breaking of the electroweak symmetry. The conceptual difficulty in the conformal model building is the nature of the symmetry, which is sometimes misleadingly called classical scale invariance. This symmetry is anomalous, since generically the renormalization-group (RG) running  of the parameters leads to a non-vanishing trace of the energy-momentum tensor~(EMT), which enters the divergence of the scale current. 

We now argue that, if conformal invariance is a fundamental symmetry of Nature, then the quantum field theory must have a vanishing trace anomaly at some scale. 
In the absence of explicit mass parameters, the trace of the EMT is given by a weighted sum of the beta functions. The anomalous Ward identity thus allows only logarithmic dependence of physical quantities on the renormalization scale. 
Any quadratically divergent contributions to the Higgs mass must therefore be purely technical and are typically introduced by explicitly breaking the conformal invariance by regulators. The formal divergences can be absorbed by appropriate counterterms.

In the Standard Model, the hypercharge gauge coupling is not asymptotically free and thus will increase with energy. In this context, there are two options to still accomplish a vanishing trace anomaly. First, the SM gauge group is embedded in a non-Abelian group so that the corresponding coupling is asymptotically free \cite{Pelaggi:2015kna}. Second, the hypercharge contribution to the trace anomaly is canceled by the gravitational anomaly. This is possible as the anomalous gravitational contribution to the trace can be negative \cite{Starobinsky:1980te} given certain values of the couplings of scalar fields and the curvature scalar.
We will demonstrate how this can work in a toy model set-up.
We argue that this vanishing of the trace of the EMT is a necessary matching condition between the low-energy theory and the UV-complete conformal embedding. 
  
If the second possibility is realized, from the point of view of a low-energy theory, this means that the electroweak symmetry is broken by radiative corrections without tree-level mass parameters. Furthermore, the theory must allow a RG evolution up to the Planck scale, at which the gravitational contributions become relevant. This means in particular that in the RG evolution no Landau poles or vacuum instabilities appear below the Planck scale. Moreover, no explicit threshold scales can be located in between the Planck scale and the low-energy theory.
At this point we emphasize that the focus of the present paper lies on the physics of a conformal theory \textit{below} the Planck scale. In this energy regime the theory is described by a renormalizable quantum field theory, the radiative behavior of which is expressed in terms of the RG running. The criteria discussed in this article are necessary conditions for any extension of the Higgs sector in order to enable stable RG running up to the Planck scale. It is not the purpose of this paper to give a definite answer to the question of what is the physics beyond the Planck scale. However, we will address the question of how an effectively conformal model \textit{may} emerge from an embedding including gravity.
The gravity scale itself can be generated spontaneously, see \cite{Adler:1982ri} for a review. Of particular interest are Yang-Mills theories which can lead to a spontaneous scale of gravity in a conformal set-up. We note, however, that this process can happen without further influence on the theory below the Planck scale and gravity might emerge with an explicit scale and also induce a gravitational conformal anomaly. We will use this fact to demonstrate that the trace of the EMT can vanish at a particular scale, leading to a vanishing of the total conformal anomaly.

Our analysis changes the perspective under which the hierarchy problem is viewed. The question is not why in a given model the Higgs mass is light, but rather whether a quantum field theory with a given set of fields and parameters is stable under renormalization group translations. This RG stability will be our essential criterion to distinguish models and to analyze whether a particular parameter configuration is allowed. This criterion selects certain representations which can be added to the SM. We find that only the interplay of scalars, fermions and gauge bosons can lead to the desired RG stability.

In this paper we revisit several classically scale-invariant models and investigate whether they can be low-energy realizations of a conformal theory. Including all relevant effects we find that in contrast to previous studies, for example \cite{Meissner:2006zh}, the SM extension by one real scalar field is not consistent with this requirement. Eventually, we identify the minimal conformal extension of the SM Higgs sector to consist of the usual complex Higgs doublet supplemented by two real scalar gauge singlets, one of which develops a non-zero vacuum expectation value~(vev). In this context, minimality implies that the SM gauge group is not altered and the additional number of representations is minimal. We find that the scalar field without the vev can be a viable dark matter candidate.
Furthermore, small neutrino masses can be easily accommodated in this model.
Another important result of our work is that the physical Higgs will have sizable admixtures of one of the singlet scalars which can be used to constrain our model's parameter space.

We present our analyses in Section \ref{sec:minimalModel}. First, we describe the method used in this paper to obtain our results. After that we scan through the most simple conformal models, starting with the extension of the SM by one additional scalar. We then systematically investigate further scalar extensions until we find a successful model. 
We will discuss the matching of the low-energy theory to the semi-classical regime in gravity in Section \ref{sec:SemiclassicalGravity}.
In Section \ref{sec:implications} we discuss important implications of our findings and summarize our results.


\section{Finding the minimal model}
\label{sec:minimalModel}
\noindent
One of the central aspects in the Standard Model is the spontaneous breaking of electroweak symmetry induced by a negative mass parameter of the Higgs field.
In a conformal extension of the SM, without any explicit mass scale present at tree level, the spontaneous breakdown must be triggered by quantum effects.
The corresponding mechanism was first investigated by Coleman and E.\,Weinberg in the context of massless scalar QED \cite{Coleman1973}.
There, the authors showed that even if a theory possesses a symmetric vacuum at tree level, the one-loop effective potential may exhibit a non-trivial minimum which then induces spontaneous symmetry breaking (SSB).
In other words, radiative corrections dynamically generate a mass scale in a classically conformal model. A scale generated in this way obviously also breaks the (anomalous) conformal symmetry spontaneously. Accordingly, we expect the theory's low-energy phase to contain one pseudo-Goldstone boson (PGB) which obtains its finite mass only at loop level.
Note that from the low energy perspective the PGB discussed here can be described by an effective theory of the dilaton, for a detailed discussion on the phenomenology see \cite{Goldberger:2008zz}.

From a more technical point of view, determining the effective potential's minimum is typically a challenging task in models of several scalars. However, there exists a method due to Gildener and S.\,Weinberg which allows a systematic minimization \cite{Gildener1976b}. In their formalism, minimization conditions manifest themselves as implicit equations for the model's scalar couplings, the so-called Gildener-Weinberg conditions.
Due to the couplings' running, these conditions will only be satisfied at a particular energy, which then is to be identified with the scale of SSB, henceforth referred to as the Gildener-Weinberg scale $\GWscale$.
We review the basic principles and some technical details of the Gildener-Weinberg formalism in Appendix~\ref{sec:GWformalism}. In particular, we will introduce the loop function $B$ in \eqref{GW_formalism_LoopFunctionsAB} which will play a central role in our analysis. It quantifies the effective potential's curvature at its minimum and thus also the PGB mass squared (\cf \eqref{GW_formalism_PGBmass}). Consistency requires $B$ to be positive.

It is well known that radiative symmetry breaking \`a la Coleman-Weinberg does not work in the SM due to the large top quark mass \cite{Abazov:2004cs}. In the Gildener-Weinberg formalism this failure is reflected in the fact that $B = \BSM$ is negative such that the effective potential does not develop a minimum but a maximum. In order to render $B$ positive, one has to achieve a dominance of bosonic degrees of freedom (see \eqref{GW_formalism_LoopFunctionsAB}). By this line of argumentation, it is immediately clear that no model can work in which the SM is extended by fermionic representations only. In particular, the SM supplemented by right-handed neutrinos cannot facilitate radiative SSB. Hence, it is necessary to add bosonic degrees of freedom to the theory.

The question for the rest of this work will be: What is the \textit{minimal} configuration to enable radiative SSB with successful RG running up to the Planck scale? In this context minimality implies that the SM gauge group, 
\begin{align*}
	\text{SU(3)}_c \times \text{SU(2)}_L \times \text{U(1)}_Y \fineq{,}
\end{align*}
is not altered and the additional number of representations is minimal. If two models are equal according to the above criteria, the number of parameters selects the minimal model. In particular, we will not add any new \textit{gauge} degrees of freedom. Note that the scalar degrees of freedom added to the model in principle may or may not acquire finite vacuum expectation values, depending on their quantum numbers. 

In the models under investigation we find that obtaining scalar couplings, which allow for a successful RG running up to the Planck mass $\PlanckScale$, turns out to be a tightrope walk.
On the one hand, the couplings need to be large enough at the GW scale in order to have sufficiently heavy new scalars which then render $B$ positive at low energies.
On the other hand, starting with too large scalar couplings at $\GWscale$ will inevitably lead to low-scale Landau poles in the scalar sector.

The method used in our analysis is as follows. First, we choose the class of models we want to investigate. Then, we derive the corresponding potential and the one-loop beta functions. The unknown scalar couplings introduced by the potential constitute our initial parameter space. We use the Gildener-Weinberg formalism to obtain the theory's vacuum and from this derive the masses of the physical scalar modes.

In doing so, we ascertain that the well-established physics of electroweak symmetry breaking (EWSB) is preserved. For instance, we will directly exclude models which imply a significant shift of the $\rho$-parameter. We then explicitly check whether the observed values for the Higgs mass $\mHiggs$ and the electroweak scale $v$ are properly reproduced.
As an additional consistency requirement, we make sure that the scalar to be identified with the Higgs boson observed at the LHC, $\HiggsField$, mainly consists of the field that couples to the SM fermions. An experimental bound on the mixing of $\HiggsField$ to other scalars is given by $\absVal{\sin \beta} \leq \num{0.44}$ \cite{Farzinnia:2013pga,Farzinnia2014}.
Together with appropriate Gildener-Weinberg conditions, all of the above constrains will allow us to limit the model's parameter space and obtain initial conditions for the renormalization group equations (RGEs).

In a first analysis of a given model's RG running, we apply a \enquote{best-case approximation}. Thus, we obtain a conservative estimate for the largest possible scale $\UVscale$ at which at latest an instability occurs or the theory's couplings turn non-perturbative.
If the scale found in this way is significantly smaller than the Planck mass, we exclude the model in accordance with our previous discussion.
Otherwise, we perform a numerically more challenging but completely consistent calculation in order to determine the actual value for $\UVscale$.

In the following we are going to present the results of our study. In Section \ref{sec:oneScalar} we discuss the simplest extension of the SM by one additional scalar. We analyze the next-to-simplest case of adding two scalar degrees of freedom in Section \ref{sec:twoScalars}. This set-up contains the minimal extension of the SM that leads to correct SSB and successful RG running up to the Planck scale.


\subsection{SM + one scalar representation} \label{sec:oneScalar}
\noindent
In accordance with our discussion in the previous paragraph, the simplest extension of the SM which might allow for radiative symmetry breaking is obtained by adding a scalar gauge singlet. Generalizing this ansatz, we investigate models in which one in general complex, colorless scalar SU(2)$_L$ multiplet with given hypercharge is added to the SM,
\begin{align}
	\chi \sim (1, \, N, \, Y) \fineq{.}
	\label{eq:oneScalar:QNs}
\end{align}
The scalar potential consistent with the SM gauge symmetries and scale invariance, which our discussion will be based on, reads as follows
\begin{align}
	\begin{split}
		V ={}& \lambda_1(\phi^\dagger \phi)^2 + \lambda_2 (\chi^\dagger\chi)^2
		+ \lambda_3 (\chi^\dagger T^a \chi)^2 \\
		& + \kappa_1 (\phi^\dagger\phi) (\chi^\dagger \chi)
		+ \kappa_2 (\phi^\dagger \tau^a \phi) (\chi^\dagger T^a \chi) \fineq{,}
	\end{split}
	\label{eq:oneScalar:general_potential}
\end{align}
where $\phi=(\phi^+, \, \phi^0)^\transpose$ denotes the usual complex Higgs doublet and $T^a$ are the generators of the SU(2) Lie algebra in the $N$-dimensional irreducible representation (irrep) under which $\chi$ transforms. Accordingly, $\tau^a$ denote the generators of SU(2) in the fundamental representation. 

Note that further gauge-invariant operators of the form $(\chi^\dagger T^{a_1}\ldots T^{a_n}\chi)^2$ as well as the corresponding portal terms are, in principle, present in the potential \eqref{eq:oneScalar:general_potential}.\footnote{Depending on the dimension $N$, some operators might be redundant in the sense that they can be expressed as a linear combination of operators containing less generator matrices. Accordingly, no additional coupling is introduced in those instances.} However, the authors of \cite{Hamada2015c} have found that for the most stable RG running all associated couplings have to vanish in the infrared, \ie at the Gildener-Weinberg scale in our context. Nevertheless, we include the simplest representatives of the above operators, namely the $\lambda_3$ and $\kappa_2$ term.

Besides, there exist additional operators, which are only invariant for special combinations of $N$ and $Y$. Again motivated by the results of \cite{Hamada2015c}, we will in general ignore those terms. In cases in which we take them into account, we will discuss them separately.

Checking the consistency of the models of interest necessarily requires knowledge about the corresponding RGEs. Therefore, we have calculated the one-loop beta functions for these models and list the results in Appendix \ref{app:rge}. Before we investigate the most general case, let us first restrict the discussion to the situation in which $\chi$ represents a \textit{real} multiplet.


\subsubsection{Real multiplet with zero vacuum expectation value}
\label{sec:oneScalar:real-woVEV}
\noindent
Let $\chi$ for the moment be a real SU(2)$_L$ multiplet in the sense that it coincides with its charge conjugate field, \ie
\begin{align}
	\tilde{\chi} := \mathcal{C} \chi^* \stackrel{!}{=} \chi \fineq{,}
	\label{eq:oneScalar:realMultiplet}
\end{align}
where $\mathcal{C}$ is a suitable charge conjugation matrix.\footnote{For the proper definition of $\mathcal{C}$, see the discussion after \eqref{eq:oneScalar:kappa_3}.} As obvious from the above definition, real multiplets necessarily have zero hypercharge. Furthermore, it is easy to show that the term $\chi^\dagger T^a \chi$ vanishes identically for all real fields transforming under an arbitrary irrep of SU(2)$_L$. Hence, the only non-zero terms in the general potential \eqref{eq:oneScalar:general_potential} are those proportional to $\lambda_1, \lambda_2$ and $\kappa_1$. The potential therefore reduces to
\begin{align}
	V = \lambda_1 (\phi^\dagger \phi)^2 + \lambda_2 (\chi^\dagger \chi)^2  + \kappa_1 (\phi^\dagger \phi)(\chi^\dagger \chi) \fineq{.}
	\label{eq:oneScalar:ON_potential}
\end{align}
Notice that the above potential enjoys an accidental global $O(4)\times O(N)$ symmetry. Since only the odd-dimensional irreps of SU(2) are real (as opposed to pseudo-real), multiplets satisfying the reality condition in \eqref{eq:oneScalar:realMultiplet} can only be consistently defined for odd $N$.


For real scalar multiplets $\chi$, always one of its component fields is electrically neutral and may therefore acquire a finite vev. We will discuss this case separately later. For now, let us assume that all $\chi_i$ have zero vacuum expectation value. Then the electroweak vev is just that of the Higgs doublet, $v\equiv v_\phi$, and the new scalar's component fields \textit{all} obtain a finite mass during EWSB ($\phi^0 = v + h/\sqrt{2}$),
\begin{align}
	m_\chi^2 = 2 \kappa_1 v^2 \fineq{.}
	\label{eq:oneScalar:chiMass-woVEV}
\end{align}
Similarly, the mass of the physical Higgs mode $\HiggsField \equiv h$ is given by $\mHiggs^2 = 6\lambda_1 v^2$ at tree level. Since all physical masses have to be real, the above formula shows that the portal coupling $\kappa_1$ is necessarily non-negative at the GW scale.

Next, the Gildener-Weinberg condition corresponding to the assumed vev configuration is $\lambda_1(\GWscale) = 0$. Accordingly, the tree-level mass of the Higgs vanishes at $\GWscale$, which implies that the physical Higgs is to be identified with the PGB of broken scale invariance. Hence, working in the GW formalism, the physical Higgs mass at $\GWscale$ is to be calculated via the one-loop formula given in \eqref{GW_formalism_PGBmass}, \ie
\begin{align}
	\mHiggs^2 = 8(\BSM + B_\text{add}) \chevron{\varphi}^2 = 8 B_\text{add} v^2 - K \fineq{,}
	\label{eq:oneScalar:PGBmass}
\end{align}
where $K := -8\BSM v^2 > 0$ and $\chevron{\varphi}=v$ denotes the condensate introduced after \eqref{GW_formalism_FlatDirection_Definition} in the Appendix. Furthermore, it follows from \eqref{GW_formalism_LoopFunctionsAB} and \eqref{eq:oneScalar:chiMass-woVEV} that
\begin{align}
	B_\text{add} = \frac{N \kappa_1^2}{16\pi^2} \fineq{.}
	\label{eq:oneScalar:Badd}
\end{align}
\eqref{eq:oneScalar:PGBmass} and \eqref{eq:oneScalar:Badd} can now be solved for the unique portal coupling at the GW scale which is consistent with the experimental values for $\mHiggs$ and $v$
\begin{align}
	\kappa_1(\GWscale) = \frac{\pi}{v} \sqrt{2 \bigl( \mHiggs^2 + K(\GWscale) \bigr)} \cdot N^{-\nicefrac{1}{2}} \fineq{.}
	\label{eq:oneScalar:portalExact}
\end{align}
Even though $\mHiggs$ in the above equation is evaluated at $\GWscale$, we can still insert the measured value of the Higgs pole mass $\mHiggs = \SI{125}{GeV}$, since it runs logarithmically and we always assume $\ln(\GWscale) \sim \ln(\mHiggs)$.
Equation (\ref{eq:oneScalar:portalExact}) now shows that increasing the number $N$ of new scalar degrees of freedom implies a smaller value for the portal coupling for otherwise fixed quantities. In other words, introducing a large scalar multiplet helps to maintain the necessary condition $B>0$, while at the same time allowing small portal couplings. One might therefore think that for large enough $N$ Landau poles in the scalar sector can be entirely evaded.
However, $N$ also unavoidably enters in some terms of the model's beta functions leading to a faster RG running such that even for small couplings at $\GWscale$ low-scale Landau poles are possible (see also Appendix \ref{app:rge}). Since $N$ enters the problem in a non-trivial way, only an explicit calculation of the RG running can shed light on the question of whether some Landau pole exists below $\PlanckScale$ for given $N$.

In order to simplify such a calculation in the class of models under consideration, we neglect the SM contribution to the Higgs mass $K \equiv K(\GWscale)$ and set
\begin{align}
	\kappa_1(\GWscale) := \sqrt{2}\pi \frac{\mHiggs}{v} N^{-\nicefrac{1}{2}} \fineq{.}
	\label{eq:oneScalar:bca}
\end{align}
As $K$ is positive, this definition exemplifies the \enquote{best-case approximation} in the sense that for given $N$, the exact value for $\kappa_1(\GWscale)$ will always be larger than that defined in \eqref{eq:oneScalar:bca}. But the larger the initial portal coupling, the sooner one of the scalar couplings will develop a Landau pole.

Uniquely solving the given model's RGEs requires to additionally fix the value of the second quartic coupling at the GW scale $\lambda_2(\GWscale)$ as well as the renormalization point $\GWscale$ itself.
Note that setting the portal coupling according to \eqref{eq:oneScalar:bca} only guarantees the proper ratio $\mHiggs/v$, but not the correct overall scale. In a full calculation the latter, would have to be set by adjusting $\GWscale$ appropriately.
For the following study, we will, however, ignore this additional constraint and choose $\GWscale = \SI{500}{GeV}$. Since we expect the exact value of $\GWscale$ to be of the same order as $v$ and the running is not very sensitive on where we precisely start in the range $[\SI{100}{GeV},\,\SI{1}{TeV}]$, this approximation will not significantly affect the position of Landau poles.

\begin{figure}[t]
	\centering
	\includegraphics[width=0.97\columnwidth]{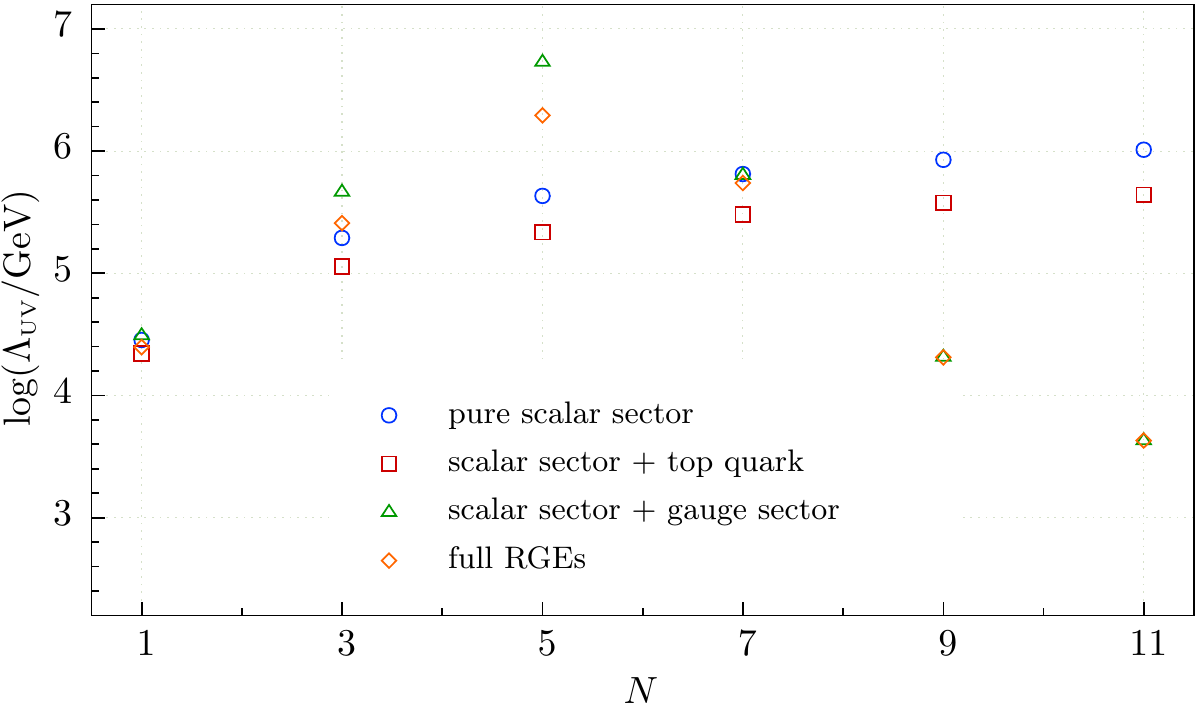}
	\caption{Largest possible UV scale in extensions of the conformal SM by one real SU(2)$_L$ $N$-plet with vanishing vev. The color code indicates which set of beta functions and couplings are taken into account.}
	\label{fig:oneScalar:UVscale-woVEV}
\end{figure}

Lastly, we vary $\lambda_2(\GWscale)$ in the perturbative range and eventually employ the value which allows the farthest extrapolation into the UV. For given order $N$, Figure \ref{fig:oneScalar:UVscale-woVEV} shows the largest possible scale $\UVscale$ at which at least one of the model's couplings becomes non-perturbative. According to our discussion right after \eqref{eq:oneScalar:bca}, the plotted values for $\UVscale$ are to be seen as an upper bound for the true values, which is sufficient to exclude running up to the Planck scale.

\begin{figure}[b]
	\centering
	\includegraphics[width=0.97\columnwidth]{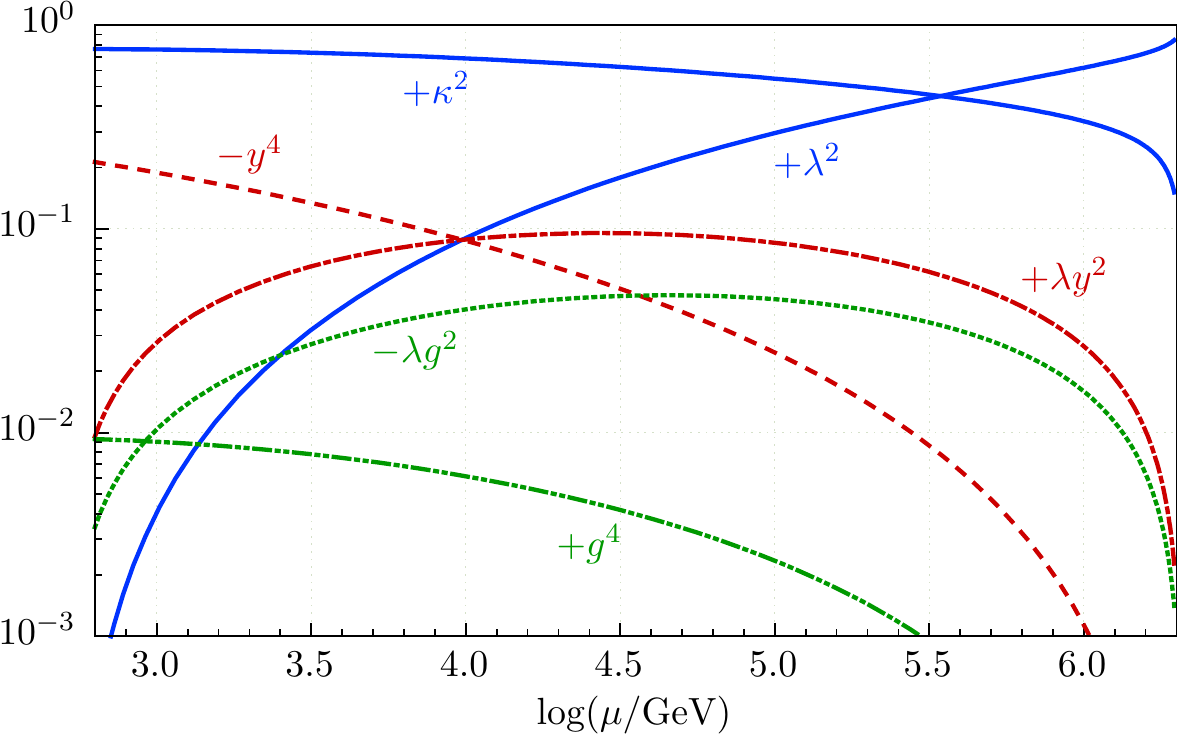}
	\caption{Running of the \textit{relative} contributions to the beta function of the Higgs self-coupling. The different contributions from the scalar, Yukawa and gauge sectors are displayed in blue, red and green, respectively. Note that the dashed red (dotted green) curve shows the absolute value of the negative contribution proportional to $- y^4$ ($-\lambda g^2$) for better comparison with the positive ones.}
	\label{fig:oneScalar:betaFctContrib}
\end{figure}

The pure scalar contribution (blue circles) supports running only up to $\log (\UVscale / \text{GeV}) \approx 6$, which is $13$ orders of magnitude below the Planck scale. This poor performance can be explained as follows:
With the scalar couplings alone, see \eqref{eq:rge:ONscalarRGEs}, no cancellation can take place and the couplings will always increase quickly. The larger the initial values of the scalar couplings the more drastic this effect becomes.
Including the top contribution into our calculation (red squares)  makes the running even worse.

To understand the effects of including the gauge sector (green triangles) we consult \eqref{eq:rge:gaugeContrib} and \eqref{eq:rge:g2}. On the one hand, the scalar beta functions receive stabilizing (negative) contributions proportional to the gauge coupling $g_2$ which grow as $N^2$. The Landau poles in the scalar RGEs are thus shifted towards larger energies for increasing $N$. Accordingly, we observe a rise in $\UVscale$ for $N\leq 5$ in Figure~\ref{fig:oneScalar:UVscale-woVEV}. On the other hand, the coefficient in the $g_2$ beta function becomes positive for large enough $N$ and a Landau pole emerges at ever smaller energies. At some point the gauge-sector Landau pole drops below that of the scalar subsystem and thus becomes the restricting one. Correspondingly, the UV scale declines for $N>5$.

The results obtained via the full running (orange diamonds) shows that we cannot reach the Planck scale in this set-up, so that the present class of models must be discarded.

We can further illuminate the above observations by analyzing the interplay between the different contributions to the beta function of the Higgs self-coupling $\lambda_1$, illustrated in Figure~\ref{fig:oneScalar:betaFctContrib} (\cf also Appendix~\ref{app:rge:SM+real}). It shows the running of the contributions from the scalar (blue), Yukawa (red) and gauge (green) sectors with respect to the renormalization scale $\mu$. Note the logarithmic scale of the $y$-axis.
While in the gauge sector the stabilizing negative contribution $- \lambda g^2$ soon dominates over the positive $+ g^4$, it is still overpowered by the contribution $+ \lambda y^2$ which dominates the Yukawa sector for large enough scales $\mu$. However, the most important observation from Figure~\ref{fig:oneScalar:betaFctContrib} is that the relative contribution of the portal coupling $\kappa_1$ is about one order of magnitude larger than the non-scalar ones.
Correspondingly, the divergence in $\lambda_1$ is triggered by the portal term which therefore must be kept sufficiently small in order to avoid any Landau pole. However, \eqref{eq:oneScalar:portalExact} prevents small initial values for $\kappa_1$ in the present case.
Additionally, Figure \ref{fig:oneScalar:betaFctContrib} explicitly demonstrates that there is no possibility for complete cancellations between the Yukawa and gauge sectors on the one hand, and the scalar sector on the other hand. Consequently, stabilizing cancellation must spring from negative contributions within the scalar sector itself.


\subsubsection{Real multiplet with finite vacuum expectation value}
\label{sec:oneScalar:real-wVEV}
\noindent
Starting again from the scalar potential in \eqref{eq:oneScalar:ON_potential}, we will now investigate the situation in which one component of the real scalar multiplet, say $\chi_{m_0}$, develops a \textit{finite} vev, \ie $\chi_{m_0} = v_\chi + \sigma$. Since the symmetry group of electromagnetism is observed to be unbroken at low energies, only electrically neutral components of $\chi$ may acquire a non-zero vev. Together with $Y=0$ for real multiplets this gives the relation $m_0 = (N + 1)/2$.

Anticipating the common origin of both vevs in the Coleman-Weinberg mechanism and adopting the notation of Appendix \ref{sec:GWformalism}, we parametrize
\begin{align}
	\begin{split}
		v_\phi & = n_1 \chevron{\varphi} \equiv \sin \alpha \chevron{\varphi} \fineq{,}\\
		v_\chi & = n_2 \chevron{\varphi} \equiv \cos \alpha \chevron{\varphi} \fineq{,}
	\end{split}
	\label{eq:oneScalar:vevAlignment1}
\end{align}
thereby defining the vev alignment angle $\alpha \in (0,\,\pi/2)$.
Following the steps of Appendix \ref{sec:GWformalism}, it is now straightforward to write down the GW conditions for the model under consideration and deduce the following identities which define the energy scale of spontaneous symmetry breaking $\GWscale$
\begin{align}
	4\lambda_1 \lambda_2 - \kappa_1^2 = 0 \sepeq{,} n_1^2 = \frac{\kappa_1}{\kappa_1 - 2 \lambda_1} \fineq{.}
	\label{eq:oneScalar:GWconditions}
\end{align}
We emphasize that all couplings in the above relations are to be understood as evaluated at $\GWscale$. Furthermore, we have $n_2^2 = 1-n_1^2$ and we see that $n_1^2$ can only be between zero and one for positive $\lambda_1$ if the portal coupling is negative.
Combining \eqref{eq:oneScalar:vevAlignment1} and \eqref{eq:oneScalar:GWconditions}, we find the vev alignment angle in terms of scalar couplings at $\GWscale$, namely
\begin{align}
	\tan\alpha \equiv \frac{v_\phi}{v_\chi} = \sqrt{-\frac{\kappa_1}{2\lambda_1}} \fineq{.}
	\label{eq:oneScalar:vevAlignment2}
\end{align}
We will use this formula in a moment to obtain information about the relative magnitude of the two vevs.

First, however, let us remark that for finite $v_\chi$, the CP-even degrees of freedom $\sigma$ and $h$ will in general mix and it is not clear a priori which mass eigenstate is to be identified with the physical Higgs boson found at the LHC. To answer this question, we consider the scalar mass matrix of the neutral, CP-even modes $\tilde{\Phi} = (h, \sigma)^\transpose$ defined via $V \supseteq \tfrac{1}{2} \tilde{\Phi}^\transpose \tilde{M}^2 \tilde{\Phi}$, which can be computed from the potential in \eqref{eq:oneScalar:ON_potential} as
\begin{align}
	\tilde{M}^2 =
	\begin{pmatrix}
		6\lambda_1 v_\phi^2 + \kappa_1 v_\chi^2 & 2\sqrt{2} \kappa_1 v_\phi v_\chi \\
		2\sqrt{2} \kappa_1 v_\phi v_\chi & 12 \lambda_2 v_\chi^2 + 2 \kappa_1 v_\phi^2
	\end{pmatrix}
	\fineq{.}
	\label{eq:oneScalar:massMatrix}
\end{align}
Since $\tilde{M}^2$ is symmetric and real it can be diagonalized by an orthogonal matrix $U$, conveniently parametrized by a single mixing angle $\beta$. The two mass eigenstates $(\Phi_1,\Phi_2)^\transpose = U\tilde{\Phi}$ can then be written as
\begin{align}
	\begin{split}
		\Phi_1 = \cos \beta \cdot h - \sin \beta \cdot \sigma \fineq{,} \\
		\Phi_2 = \sin \beta \cdot h + \cos \beta \cdot \sigma \fineq{,}
	\end{split}
	\label{eq:oneScalar:massEigenstates}
\end{align}
one of which will have to be identified with the physical Higgs boson $\HiggsField$. To see which one, note that we can assume $\beta \in (-\pi/4,\,\pi/4)$ without loss of generality. But then we immediately see the necessity of $\Phi_1 \equiv \HiggsField$, since low-energy phenomenology requires the Higgs state to consist mainly of the SM doublet field \cite{Farzinnia:2013pga,Farzinnia2014}.

The mass matrix in \eqref{eq:oneScalar:massMatrix} has two distinct mass eigenvalues, which are given by
\begin{align*}
	2 m_\pm^2 = \tr(\tilde{M}^2) \pm \sqrt{\bigl[ \tr(\tilde{M}^2) \bigr]^2 - 4 \cdot \det (\tilde{M}^2)} \fineq{.}
\end{align*}
As before, we can exploit the additional relations between the scalar couplings given in \eqref{eq:oneScalar:GWconditions} together with the constraints $\lambda_1>0$ and $\kappa_1<0$ to obtain expressions for the above tree-level masses at $\GWscale$, namely
\begin{align}
	m_+^2 = 4(\lambda_1 - \kappa_1) v_\phi^2 \sepeq{,}
	m_-^2 = 0 \fineq{.}
	\label{eq:oneScalar:massEigenvaluesGW}
\end{align}
As expected, the spectrum in the broken phase still contains one scalar degree of freedom with vanishing tree-level mass, the PGB of broken scale invariance. In contrast, $m_+^2$ is always positive. Which of the mass eigenstates $\Phi_i$ is now to be identified with the PGB depends on the sign of the scalar mixing angle. The correct assignment procedure can be deduced by simply calculating the diagonalized mass matrix for both cases. For positive $\beta$, we obtain $U \tilde{M}^2 U^\transpose = \operatorname{diag}(m_+^2, m_-^2)$ such that $\Phi_2$ is the PGB, whereas the diagonal entries are exchanged for negative $\beta$ and $\Phi_1$ corresponds to the PGB (\cf Table \ref{tab:oneScalar:twoCasesFiniteVEV}).

Next, let us derive an expression for $\beta$ in terms of model parameters by requiring the matrix $U\tilde{M}^2 U^\transpose$ to be diagonal. An explicit calculation yields
\begin{align*}
	\tan 2\beta = \frac{4\sqrt{2} \kappa_1 \tan\alpha}{(12\lambda_2 - \kappa_1) - 2(3\lambda_1 - \kappa_1) \tan^2\alpha} \fineq{,}
\end{align*}
where we used \eqref{eq:oneScalar:vevAlignment1} in order to introduce the vev alignment angle $\alpha$. The above identity shows that in a general theory the relation between the angles $\alpha$ and $\beta$ explicitly depends on the scalar couplings. In particular, experimental constraints for one angle do not directly translate into bounds for the other one, unless all involved couplings are known.
In contrast, using the additional restrictions imposed on the scalar couplings by the GW condition in \eqref{eq:oneScalar:GWconditions}, we can rewrite the above equation as
\begin{align}
	\tan 2\beta = \frac{2\sqrt{2} \tan\alpha}{1 -2 \tan^2\alpha} \fineq{,}
	\label{eq:oneScalar:angleRelationGW}
\end{align}
which has the two solutions listed in Table \ref{tab:oneScalar:twoCasesFiniteVEV}. Combining the above identity with \eqref{eq:oneScalar:vevAlignment2}, we can deduce a relation between the sign of $\beta$ and the relative magnitude of the two vevs, see once more Table \ref{tab:oneScalar:twoCasesFiniteVEV}.

\begin{table}[t]
	\centering
	\begin{tabular}{lccc}
		\toprule
		 & \symhspace{7mm}{PGB} & \symhspace{3mm}{$\tan\beta(\alpha)$} & \symhspace{3mm}{$\sqrt{2}v_\phi/v_\chi$} \\
		\colrule
		$\beta < 0$ & $=\HiggsField$ & $-(\sqrt{2}\tan\alpha)^{-1}$ & $>1$ \\
		$\beta > 0$ & $\neq\HiggsField$ & $\sqrt{2}\tan\alpha$ & $<1$ \\
		\botrule
	\end{tabular}
	\caption{Summary of differences between positive and negative scalar mixing angle $\beta\in(-\pi/4,\,\pi/4)$. The assignment in the first column is done according to the discussion right after \eqref{eq:oneScalar:massEigenvaluesGW}. The statements in the second (third) column follow from \eqref{eq:oneScalar:angleRelationGW} (and \eqref{eq:oneScalar:vevAlignment2}).}
	\label{tab:oneScalar:twoCasesFiniteVEV}
\end{table}

In order to see whether we can construct a consistent conformal model in the present set-up, let us now study the two cases $\beta>0$ and $\beta<0$ separately, starting with the former. From Table \ref{tab:oneScalar:twoCasesFiniteVEV} we learn that for positive mixing angle the vev of the additional SU(2)$_L$ multiplet is sizable, namely $v_\chi > \sqrt{2}v_\phi$. The presence of such a vev will in general significantly shift away the $\rho$-parameter from its experimentally well-established SM-like value of $\rho \approx 1$ \cite{Agashe:2014kda}. However, there are exceptions to this. Considering real multiplets, it is only the singlet which does not affect the $\rho$-parameter. Hence, for positive $\beta$ we can restrict the discussion of the additional real scalar with vev to this case. 

Furthermore, a positive mixing angle implies $m_+ = \mHiggs$, \ie the physical Higgs \textit{cannot} be identified with the PGB. Consequently, only the Higgs mass contributes to $B_\text{add}$ in \eqref{eq:oneScalar:PGBmass}. But obviously, the LHC Higgs is not heavy enough to compensate the large, negative top quark contribution to $B$ and the PGB therefore obtains a negative mass-square. In other words, the one-loop effective potential exhibits a maximum instead of a minimum at the electroweak scale which is clearly unphysical and rules out this scenario.

Moving to negative scalar mixing angles, we now have $\sqrt{2} v_\phi > v_\chi$ (\cf Table \ref{tab:oneScalar:twoCasesFiniteVEV}). So a priori $v_\chi \ll v_\phi$ is allowed and the additional vev's contribution to the $\rho$-parameter can in principle be sufficiently small.
For $\beta<0$, we will therefore not only investigate the singlet case, but also larger multiplets.
Note at this point that for $N>1$ a non-zero vev in the $\chi$-sector spontaneously breaks $O(N)\longrightarrow O(N-1)$. The theory's spectrum in the broken low-energy phase will thus contain $N-1$ Goldstone modes. Consequently, only one component field of $\chi$ will acquire a non-vanishing mass term.

Furthermore, negative $\beta$ implies that the physical Higgs is to be identified with the PGB (\cf Table \ref{tab:oneScalar:twoCasesFiniteVEV}) and the theory's spectrum contains one additional scalar with unknown mass $m_+$. Correspondingly, we can calculate
\begin{align*}
	B_\text{add} = \frac{n_1^4}{4\pi^2} (\lambda_1-\kappa_1)^2
\end{align*}
and use \eqref{eq:oneScalar:PGBmass} to eventually arrive at the constraint
\begin{align}
	(\SI{125}{GeV})^2 \stackrel{!}{=}
	\mHiggs^2 = \frac{2}{\pi^2} n_1^2 (\lambda_1 - \kappa_1)^2 v_\phi^2 - K \fineq{.}
	\label{eq:oneScalar:PGBmass_betaNegative}
\end{align}
The electroweak vev can generically be written as $v^2 = \left( 2 \sqrt{2} G_F \right)^{-1} = v_\phi^2 + c v_\chi^2$, with a model-dependent non-negative constant $c$, implying $v_\phi\leq v$. Hence, we can rewrite \eqref{eq:oneScalar:PGBmass_betaNegative} as a condition on the unknown couplings evaluated at $\GWscale$
\begin{align*}
	n_1^2 (\lambda_1 - \kappa_1)^2 \stackrel{!}{>} \pi^2 \, \frac{\mHiggs^2}{2 v^2}
	=: r^2 \fineq{,}
\end{align*}
which now depends on the empirically known quantities $v$ and $\mHiggs$. We have used again that $K$ is positive. Replacing the above inequality by an equality corresponds to the \enquote{best-case approximation} in a similar sense as discussed right below \eqref{eq:oneScalar:bca}. Solving \eg for $\lambda_1$, one obtains
\begin{align}
	\lambda_1(\kappa_1) = \frac{\bigl( r^2 - \kappa_1^2 \bigr) \pm \sqrt{r^2 \bigl( r^2 - \kappa_1^2 \bigr)}}{-\kappa_1} \fineq{,}
	\label{eq:oneScalar:1D-solutionMF}
\end{align}
where only the solution with the plus sign gives positive $\lambda_1$ in the relevant $\kappa_1$-range (small and negative).
Furthermore, general arguments allow to constrain the valid range for the portal coupling. Firstly, $\lambda_1$ is assumed to be real, which directly gives $\absVal{\kappa_1} \leq r$. Secondly, as a consequence of \eqref{eq:oneScalar:vevAlignment2}, negative scalar mixing angles imply $\lambda_1 \leq -\kappa_1$. This, in turn, is only satisfied for $\absVal{\kappa_1} \geq \sqrt{3}r/2$. Using the numerical value for $r$, we can constrain $\kappa_1$ to
\begin{align}
	1.38 \leq \absVal{\kappa_1} \leq 1.60 \fineq{.}
	\label{eq:oneScalar:allowedRange-real-wVEV}
\end{align}
For the following study, we will choose again $\GWscale=\SI{500}{GeV}$ and vary $\kappa_1(\GWscale)$ in the allowed range. The remaining initial conditions $\lambda_1(\GWscale)$ and $\lambda_2(\GWscale)$ are then fixed by \eqref{eq:oneScalar:1D-solutionMF} and \eqref{eq:oneScalar:GWconditions}, respectively.

\begin{figure}[t]
	\centering
	\includegraphics[width=0.97\columnwidth]{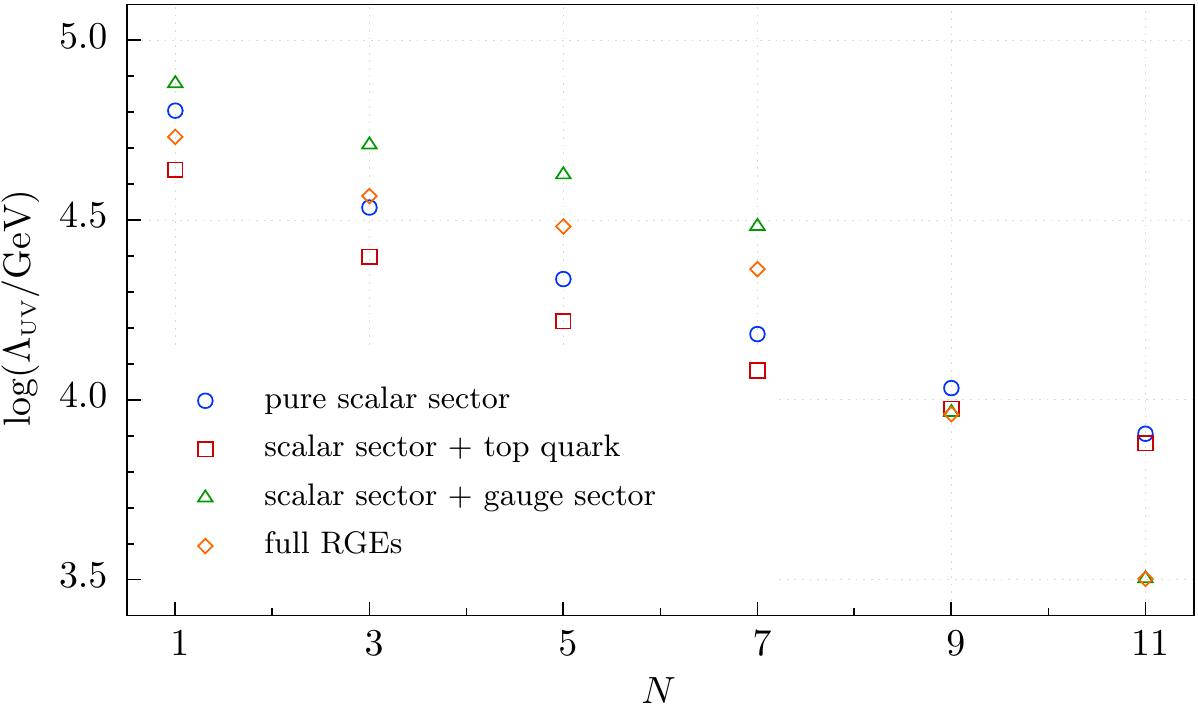}
	\caption{Largest possible UV scale in extensions of the conformal SM by one real SU(2)$_L$ $N$-plet with finite vev $v_\chi < \sqrt{2} v_\phi$. The color code indicates which set of beta functions and couplings are taken into account.}
	\label{fig:oneScalar:UVscale-wVEV}
\end{figure}

In complete analogy to Figure \ref{fig:oneScalar:UVscale-woVEV} from the last subsection, Figure \ref{fig:oneScalar:UVscale-wVEV} now illustrates the results for the present case: The largest possible scale $\UVscale$ at which at least one of the model's couplings develops a Landau pole is plotted as a function of $N$, the dimension of the additional scalar SU(2)$_L$ representation. The most important result lies in the fact that also within the present class of models, there is no representative which allows an extrapolation all the way up to $\PlanckScale$. Except for the singlet case ($N=1$), all models develop a Landau pole at even lower scales compared to the corresponding case without vev. 

The relative magnitudes of the calculated UV scales for different sets of beta functions is very similar to those observed in Figure \ref{fig:oneScalar:UVscale-woVEV} and the discussion there can be adopted.
Nevertheless, there are some qualitative differences between the two set-ups. Whereas, for instance, Figure \ref{fig:oneScalar:UVscale-woVEV} exhibits a peak for $N = 5$ in the case without vev, Figure \ref{fig:oneScalar:UVscale-wVEV} shows a strict decrease of $\UVscale$ with $N$.
This behavior can be easily understood as follows: \eqref{eq:oneScalar:1D-solutionMF}, which fixes the valid initial parameter values in the present case, does not depend on the number of added scalar degrees of freedom. Hence, for each $N$, the RG running starts from the same hypersurface in parameter space. The RGEs, however, explicitly depend on $N$ and especially the scalar contributions tend to increasingly destabilize the running for increasing $N$ (\cf \eqref{eq:rge:ONscalarRGEs}).
In the situation without vev the initial hypersurface is determined by \eqref{eq:oneScalar:bca} showing that the initial value for the portal coupling decreases with $N$. This can compensate the aforementioned destabilization for sufficiently low $N$. 

Our findings generalize the analysis of \cite{Foot:2007as} and are consistent with the conclusions of Foot et al. This concludes our discussion of extensions of the conformal SM by one \textit{real} scalar multiplet. Since we have not found a consistent theory up to now, we move on to the next class of models.

\subsubsection{Complex multiplet with zero vacuum expectation value}
\label{sec:oneScalar:complex-woVEV}
\noindent
In this section we drop the requirement of the additional scalar multiplet $\chi$ being real. Correspondingly, all calculations will be based on the potential introduced in \eqref{eq:oneScalar:general_potential} and we can drop the restriction to only odd-dimensional SU(2)$_L$ multiplets. Note that a complex scalar, as opposed to a real one, can carry non-zero hypercharge. If appropriate quantum numbers are assigned to $\chi$, one of the scalar's components can be electrically neutral and may therefore acquire a finite vev. We will discuss this case separately later.

After \eqref{eq:oneScalar:general_potential} we argued that there exist additional operators for special configurations of $N$ and $Y$, but that it is reasonable to ignore them. For the analysis of the present class of models, however, we decided to include the special term
\begin{equation}
	\label{eq:oneScalar:kappa_3}
	\Delta V_1  = \kappa_3 \Bigl[ (\phi^\transpose \varepsilon \tau^a \phi) (\chi^\transpose \mathcal{C} T^a \chi) + \hc \Bigr] \fineq{,}
\end{equation}
into the general potential. Here, $\mathcal{C}$ is a matrix in the SU(2) algebra satisfying the defining relation $\mathcal{C} T^a \mathcal{C}^{-1} = -{T^a}^\transpose$ for all $a$, and $\varepsilon$ is the two-dimensional representation of this matrix. This term then forms a gauge singlet for arbitrary $N$ as long as $Y=-Y_\phi=-\tfrac{1}{2}$ is fulfilled. However, since the matrices $\mathcal{C}T^a$ are anti-symmetric in all odd-dimensional irreps of SU(2), the $\kappa_3$-term is only present for even $N$. We decided to include $\Delta V_1$ into our analysis because it is gauge-invariant not only for one special configuration, but, as we have just learned, for all even-dimensional representations with a particular hypercharge. Nevertheless, it turns out that the best RG running is obtained for a value of $\kappa_3 \approx 0$. This further fortifies our assumption of choosing the special couplings close to zero. For better clarity, we will set $\kappa_3$ to zero in all formulas (of this subsection), even though we include $\Delta V_1$ in our calculation.

As in the real case, we will first consider the situation in which only the SM doublet acquires a finite vev, \ie $\phi^0 = v_\phi + h/\sqrt{2}$, implying that the physical Higgs mode $\HiggsField \equiv h$ is to be identified with the PGB. The associated Gildener-Weinberg condition is again $\lambda_1(\GWscale) = 0$ so that the physical Higgs only becomes massive through quantum effects with its one-loop mass squared given by \eqref{eq:oneScalar:PGBmass}.

For generic values of the portal couplings $\kappa_1$ and $\kappa_2$, all (complex) component fields $\chi_k$ will obtain some finite mass $m_k^2$ during EWSB. However, in contrast to the real case, the presence of the $\kappa_2$-term explicitly violates the formerly exact $O(N)$ symmetry and thus leads to a mass splitting between the individual components, which is proportional to $\kappa_2$. An explicit calculation yields
\begin{align}
	m_k^2 = \tfrac{1}{4} \bigl[ 4\kappa_1 - (N -2 k+1)\kappa_2 \bigr] v^2 \fineq{}
	\label{eq:oneScalar:cmplxChiMasses}
\end{align}
with $k\in \{1,\ldots,N\}$. The portal couplings are to be understood as evaluated at $\GWscale$. One can show that requiring real masses for all new scalar particles implies non-negative $\kappa_1$ at the Gildener-Weinberg scale.
Using \eqref{eq:oneScalar:cmplxChiMasses}, we can now compute
\begin{align*}
	B_\text{add} = 2\cdot\frac{1}{64\pi^2 \chevron{\varphi}} \sum_{k=1}^{N} m_k^4 = \frac{4 N \kappa_1^2 + D \kappa_2^2}{128\pi^2} \fineq{,}
\end{align*}
where the overall factor of two takes into account the complex nature of the component fields. The Dynkin index $D$ of the representation under which $\chi$ transforms is defined in Appendix \ref{app:rge}. Anticipating $K>0$, \eqref{eq:oneScalar:PGBmass} then implies
\begin{align}
	\sqrt{4 N \kappa_1^2 + D \kappa_2^2} > 4\pi \frac{\mHiggs}{v} \fineq{.}
	\label{eq:oneScalar:cmplx-solutionMF}
\end{align}
Similar to the previous discussions, we are only interested in an upper bound for the location of potential Landau poles and therefore employ the \enquote{best-case approximation}. Accordingly, we replace the inequality in \eqref{eq:oneScalar:cmplx-solutionMF} by an equality. Given $N$ and \eg $\kappa_2(\GWscale)$, we can then simply compute the corresponding value of $\kappa_1(\GWscale)>0$.

Uniquely solving the given model's RGEs requires to fix the remaining couplings at the GW scale, namely $\lambda_2$ and $\lambda_3$, as well as the renormalization point itself. In the following, we will assume $\GWscale = \SI{500}{GeV}$ and vary all unspecified parameters in the perturbative range.

\begin{figure}[t]
	\centering
	\includegraphics[width=0.97\columnwidth]{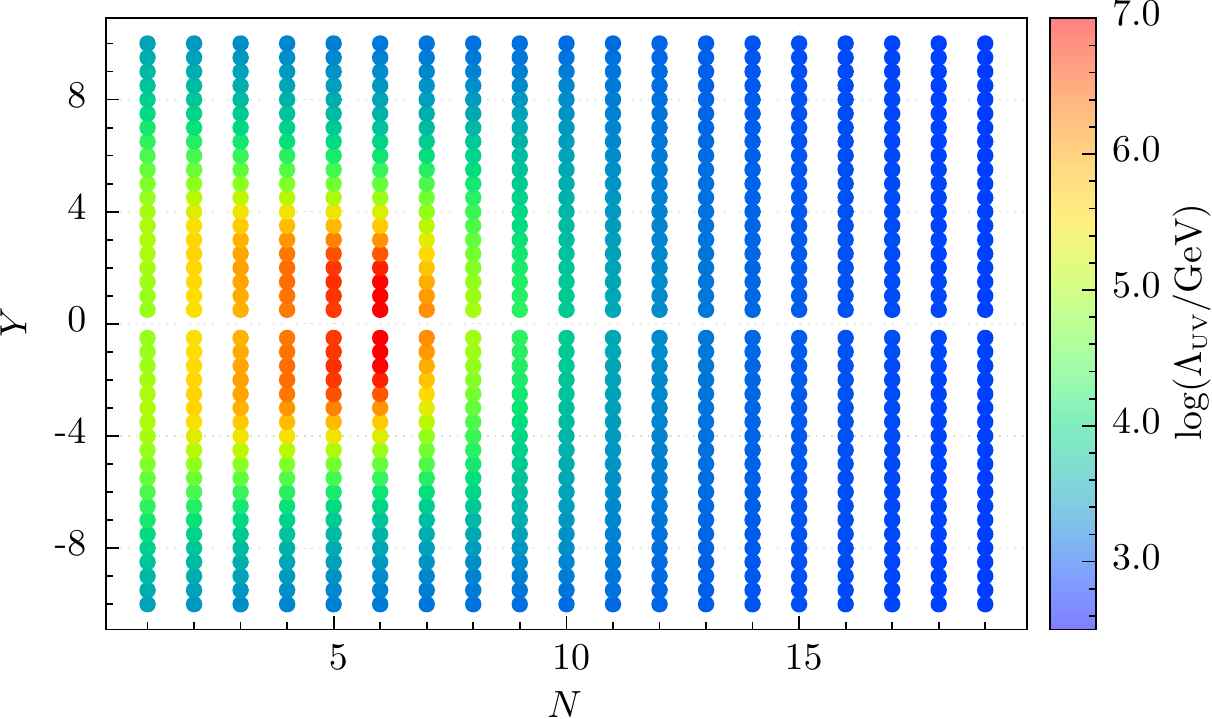}
	\caption{Largest possible UV scale in extensions of the conformal SM by one complex SU(2)$_L$ $N$-plet with hypercharge $Y$ and vanishing vev.}
	\label{fig:OneCmplxScalar:UVscale-woVEV}
\end{figure}

The results of the RG running for one additional complex representation with vanishing vev are shown in Figure~\ref{fig:OneCmplxScalar:UVscale-woVEV}. The largest possible UV scale $\log(\UVscale/\text{GeV}) \approx 7$ is obtained for $N = 5$ or $N = 6$ and small values of $Y$. The symmetry of the figure reflects the exchange symmetry of the beta functions with respect to $Y \leftrightarrow -Y$. Note that the dots for even numbers $N$ on the $Y = - \frac{1}{2}$ axis were obtained including the $\kappa_3$-term.
This term could, in principle, lead to differences in the UV scale for even-dimensional multiplets with $Y=\tfrac{1}{2}$ and $Y=-\tfrac{1}{2}$. But as we see from Figure~\ref{fig:OneCmplxScalar:UVscale-woVEV}, it has practically no effect on the RG running since the optimal initial value for $\kappa_3$ turns out to be close to zero and it is multiplicatively renormalized. Also the other additional couplings with respect to the real case, $\lambda_3$ and $\kappa_2$, are best chosen near zero at the initial scale.
Note that these findings are consistent with our discussion after \eqref{eq:oneScalar:general_potential}. Comparing Figure~\ref{fig:OneCmplxScalar:UVscale-woVEV} to the case of a real scalar without vev, Figure~\ref{fig:oneScalar:UVscale-woVEV}, we see that their features are very similar: They both support running up to about the same UV scale, which reaches its maximum at approximately the same values of $N$ and $Y$. In conclusion we find that this model, by far, does not allow for RG running up to the Planck scale.


\subsubsection{Complex multiplet with finite vacuum expectation value}
\label{sec:oneScalar:complex-wVEV}
\noindent
After discussing the case of an additional real multiplet with finite vev in Section~\ref{sec:oneScalar:real-wVEV}, the transfer to a complex representation is straightforward. First, we mention that for a complex multiplet we naturally have a different normalization for the field modes, and especially $\chi_{m_0} = v_\chi + \sigma/\sqrt{2}$ for the electrically neutral mode. As for the complex multiplet without vev, we apply the Gildener-Weinberg formalism to the general potential \eqref{eq:oneScalar:general_potential}. Introducing the following quantities 
\begin{align}
	\begin{split}
		v_\chi^\prime
		& :=
		\sqrt{2} v_\chi
		\sepeq{,}
		\tan \alpha^\prime
		:=
		\tfrac{1}{\sqrt{2}} \tan \alpha		
		\, , \\
		\lambda_2^\prime 
		&:=
		\tfrac{1}{4} \bigl[ \lambda_2 + (1 - \delta_{N,1}) \lambda_3 Y^2 \bigr] 
		\, ,
		\\
		\kappa_1^\prime  
		&:=
		\tfrac{1}{4} \bigl[ \kappa_1 + \tfrac{1}{2}(1 - \delta_{N,1})  \kappa_2 Y \bigr]
		\, ,
	\end{split}
	\label{eq:oneScalar:primedVevAndCouplings}
\end{align}
we obtain exactly the same equations as in the case with the additional \textit{real} scalar (starting from \eqref{eq:oneScalar:GWconditions}). We only need to use the primed quantities, defined in \eqref{eq:oneScalar:primedVevAndCouplings}, instead of the unprimed ones.
By this, for instance, the Gildener-Weinberg condition from \eqref{eq:oneScalar:GWconditions} now reads $4 \lambda_1 \lambda_2^\prime - \kappa_1^{\prime \, 2} = 0$. The only new aspect is an additional GW condition, namely $\kappa_3(\GWscale)=0$.
Using the aforementioned replacements, also the scalar mixing phenomenology is the same as in Section \ref{sec:oneScalar:real-wVEV}, which we summarized in Table~\ref{tab:oneScalar:twoCasesFiniteVEV}.

For positive mixing angle, Table~\ref{tab:oneScalar:twoCasesFiniteVEV} tells us that $\sqrt{2} v_\phi / v_\chi^\prime = v_\phi / v_\chi < 1$. As a consequence the additional vev is sizable and thus will in general tarnish the $\rho$-parameter. However, for $N\leq20$, there exist three complex representations which leave the $\rho$-parameter invariant, namely a singlet with $Y = 0$, a doublet with $Y = 1/2$ and a septet with $Y = 2$.
The description of one additional \textit{complex} singlet with zero hypercharge is equivalent to the description of two additional \textit{real} singlets and will be covered in Section~\ref{sec:twoScalars} (see also \eg \cite{Gabrielli2014}).
If $\chi$ is an SU(2)$_L$ doublet with $Y = 1/2$ it is a second Higgs boson and \textendash\ without additional assumptions \textendash\ would have Yukawa couplings to all of the SM fermions. This contradicts our principle of minimality and we will not further consider this case here. 
Finally, we investigated the septet model. In this case, due to the large dimensionality $N$ and the relatively large hypercharge $Y=2$, the beta function of the U(1)$_Y$ gauge coupling runs into a Landau pole before reaching the Planck scale (\cf \eqref{eq:rge:cmplxGauge}). In summary, the case of positive $\beta$ does not provide us with a consistent, minimal conformal model with a complex scalar multiplet that develops a finite vev.

For negative scalar mixing angle we have $v_\phi > v_\chi$ and for not too large $v_\chi$ the $\rho$-parameter is safe. The Higgs boson $\HiggsField$ is the PGB and its mass is generated by the additional massive scalar modes at the one-loop level. In the case of a real multiplet the potential possessed a global $O(N)$ symmetry that was spontaneously broken and, by the Goldstone theorem, guaranteed that all modes besides $\chi_{m_0}$ were massless. Here, the additional couplings $\lambda_3$ and $\kappa_2$ explicitly break this symmetry. Consequently, the masses of the charged modes are proportional to the symmetry breaking parameters.
However, as our previous analysis as well as the analysis of \cite{Hamada2015c} suggest, the couplings $\lambda_3$ and $\kappa_2$ are best chosen close to zero for optimal RG running, and the symmetry of the potential is approximately restored.
With the additional masses close to zero, the results of the model with complex multiplet acquiring a finite vev are comparable to the real case.
This argument is further substantiated by the observation \textendash\ stated within the discussion of Figure~\ref{fig:OneCmplxScalar:UVscale-woVEV} \textendash\ that the results from Section \ref{sec:oneScalar:real-woVEV} and \ref{sec:oneScalar:complex-woVEV} are both qualitatively and quantitatively very similar. Therefore, there is no reason to expect a large difference when going from real to complex $\chi$ also in the case of a finite vev.
We thus conclude that for negative $\beta$ the case of the complex multiplet with finite vev leads to similar results as in the real case shown in Figure~\ref{fig:oneScalar:UVscale-wVEV}. In particular, there will be no combination $(N,Y)$ for which the RG running can be extended far beyond $\mathcal{O}(10^7\,\text{GeV})$.

This exhausts all reasonable possibilities in the case of the conformal SM plus a complex multiplet that develops a finite vev. We have not found a consistent minimal conformal model up to now.

Before we proceed to the next class of models with two additional scalar multiplets, let us comment on the conformal SM with one additional scalar and additional \textit{fermionic} representation(s). One can easily see that this set-up will also fail to provide a consistent conformal model, since additional fermions destabilize the RG running in two ways:
Firstly, any massive fermionic particle will give negative contributions to $B_\text{add}$. In order to render $B$ positive, the scalar couplings therefore have to take larger initial values in comparison to the model without fermion.
Secondly, adding a fermion will ultimately destabilize the scalar RGEs even more due to its positive contribution proportional to $\lambda y^2$ to the beta function of a generic scalar coupling (\cf the discussion about Figures \ref{fig:oneScalar:UVscale-woVEV} and \ref{fig:oneScalar:betaFctContrib}). 
We conclude that, if a given theory develops Landau poles well below the Planck scale, then the same theory supplemented by fermions interacting via Yukawa couplings with the scalar sector will, too.


\subsection{SM + two scalar representations} \label{sec:twoScalars}
\noindent
In complete analogy to our discussion in Section \ref{sec:oneScalar}, we now consider the case in which two real scalar multiplets $\chi$ and $\xi$ are added to the SM. In doing so, we will neglect all but the standard quartic and portal couplings, in accordance with our previous analyses (\cf in particular Section \ref{sec:oneScalar:complex-woVEV}). This is also in line with our notion of minimality discussed earlier. In effect, the aforementioned assumption introduces an additional global $O(N_\chi)\times O(N_\xi)$ symmetry in the scalar sector. The associated scale-invariant tree-level potential then reads
\begin{align}
	\begin{split}
	V ={}& \lambda_\phi (\phi^\dagger \phi)^2 + \lambda_\chi (\chi^\dagger \chi)^2
	 + \lambda_\xi (\xi^\dagger \xi)^2 \\
	 & + \kappa_{\phi \chi} (\phi^\dagger \phi) (\chi^\dagger \chi)
	+ \kappa_{\phi \chi} (\phi^\dagger \phi) (\xi^\dagger \xi) \\
	& + \kappa_{\chi \xi} (\chi^\dagger \chi) (\xi^\dagger \xi) \fineq{,}
	\end{split}
	\label{eq:twoScalars:potential}
\end{align}
where $\phi$ denotes the SM complex Higgs doublet as before and both $\chi$ and $\xi$ are now supposed to satisfy reality conditions like the one in \eqref{eq:oneScalar:realMultiplet}. For the model's RGEs we again refer to the formulas given in Appendix \ref{app:rge}.

\begin{figure}[t]
	\centering
	\includegraphics[width=0.97\columnwidth]{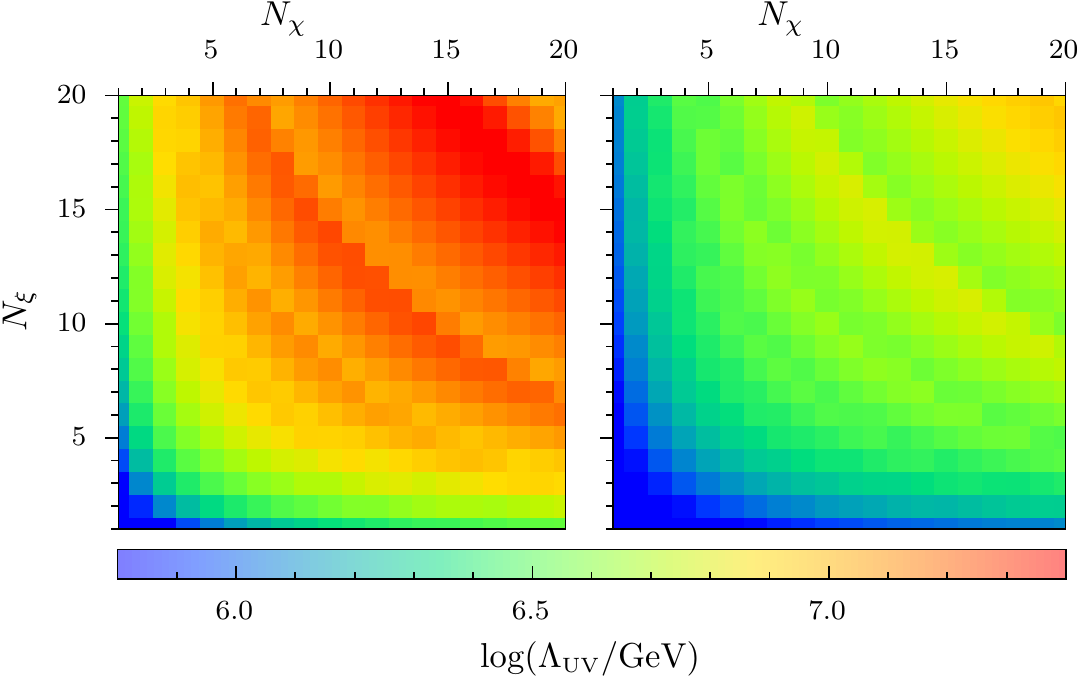}
	\caption{Largest possible UV scale in extensions of the conformal SM by two real scalar SU(2)$_L$ multiplets with vanishing vevs. The results were obtained using RGEs including \textit{Left}: scalar contributions only. \textit{Right}: scalar and top-quark contributions.}
	\label{fig:TwoScalars:UVscale-woVEV-1}
\end{figure}


\subsubsection{Two real multiplets with zero vacuum expectation value}
\noindent
Similar to our one-scalar discussion, we first assume that neither $\chi$ nor $\xi$ acquires a finite vev. In the GW formalism, this corresponds to the renormalization condition $\lambda_\phi(\GWscale)=0$. Hence, the physical Higgs is necessarily the PGB of broken scale invariance. Electroweak symmetry breaking then proceeds via $\phi^0 = v+h/\sqrt{2}$ and induces the following masses for \textit{all} new scalar degrees of freedom:
\begin{align}
	m_\chi^2 = 2 \kappa_{\phi\chi} v^2 \text{~~~or~~~}
	m_\xi^2 = 2 \kappa_{\phi\xi} v^2 \fineq{.}
	\label{eq:twoScalars:masses-woVEV}
\end{align}
Since all physical masses have to be real, the two portal couplings $\kappa_{\phi\chi}$ and $\kappa_{\phi\xi}$ are necessarily non-negative at the GW scale. The one-loop mass squared of the physical Higgs boson $\HiggsField\equiv h$ is again given by equation \eqref{eq:oneScalar:PGBmass}, but in the present situation we have
\begin{align}
	B_\text{add} = \frac{N_\chi \kappa_{\phi\chi}^2 + N_\xi \kappa_{\phi\xi}^2}{16 \pi^2} \fineq{.}
	\label{eq:twoScalars:Badd}
\end{align}
Combining the previous identity with equation \eqref{eq:oneScalar:PGBmass} and taking into account $K>0$, we arrive at
\begin{align}
	\sqrt{N_\chi \kappa_{\phi\chi}^2 + N_\xi \kappa_{\phi\xi}^2} > \sqrt{2}\pi \frac{\mHiggs}{v} \fineq{.}
	\label{eq:twoScalars:portalInequal}
\end{align}
For the purpose of finding out whether there exists a pair $(N_\chi,N_\xi)$, for which consistent radiative symmetry breaking is possible, explicit calculations of the RG running are inevitable. To facilitate those, we will again apply the \enquote{best-case approximation}, in which the above inequality \eqref{eq:twoScalars:portalInequal} is replaced by an equation. Given $N_\chi$, $N_\xi$ and \eg $\kappa_{\phi\chi}(\GWscale)$, we can then simply compute the corresponding value of $\kappa_{\phi\xi}(\GWscale)$.

Uniquely solving the given model's RGEs requires to fix the three remaining couplings at the GW scale, namely $\lambda_\chi$, $\lambda_\xi$ and $\kappa_{\chi\xi}$, as well as the renormalization point $\Lambda_\text{GW}$ itself. For the following study, we will choose $\GWscale = \SI{500}{GeV}$ and vary all unspecified couplings in the perturbative range. Whereas the quartic couplings are confined to positive values due to the requirement of vacuum stability, the sign $\kappa_{\chi\xi}(\GWscale)$ is not constrained a priori.

\begin{figure}[t]
	\centering
	\includegraphics[width=0.97\columnwidth]{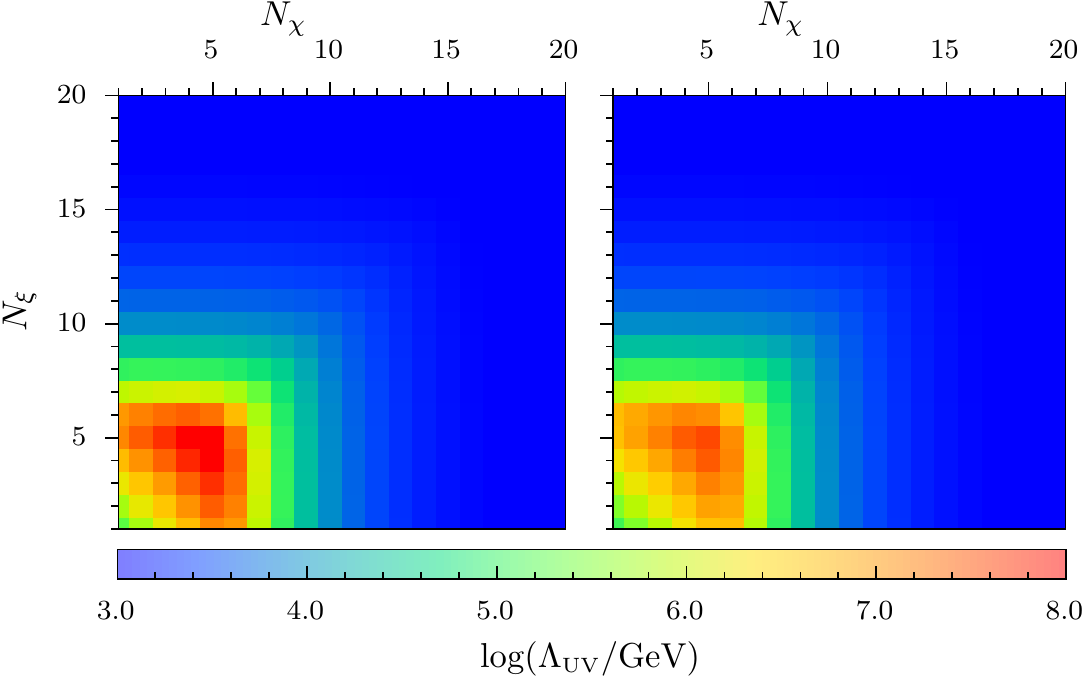}
	\caption{Largest possible UV scale in the set-up described in Figure \ref{fig:TwoScalars:UVscale-woVEV-1}. The results were obtained using \textit{Left}: RGEs including scalar and SM gauge contributions. \textit{Right}: full RGEs.}
	\label{fig:TwoScalars:UVscale-woVEV-2}
\end{figure}

Figures \ref{fig:TwoScalars:UVscale-woVEV-1} and \ref{fig:TwoScalars:UVscale-woVEV-2} summarize the findings for the largest possible UV scale $\UVscale$ we obtain working in the \enquote{best-case approximation}. In total, we show four plots, differing in the particle sectors included in the computation of the RG running. It is instructive to compare the present results to the outcome of the calculations with one extra scalar without vev (\cf Section \ref{sec:oneScalar:real-woVEV}). Thereby, each of the above panels corresponds to one set of points in Figure \ref{fig:oneScalar:UVscale-woVEV}.
On a qualitative level, each individual case gives results resembling those of its one-scalar counterpart. The respective differences between the four cases are also similar for both set-ups.
In particular, comparing the panels on the right-hand side with the ones on the left-hand side in Figures \ref{fig:TwoScalars:UVscale-woVEV-1} and \ref{fig:TwoScalars:UVscale-woVEV-2}, again exemplifies that including a generic Yukawa coupling destabilizes the flow and thus decreases the maximal possible UV scale.
On a quantitative level, the Landau poles in the present study of two additional scalars develop at somewhat higher scales compared to the corresponding divergences in the one-scalar case.

We furthermore find that for given dimensions of the scalar multiplets, the farthest RG running is obtained for vanishing quartic couplings, $\lambda_\chi$ and $\lambda_\xi$, as well as negative and often sizable $\kappa_{\chi\xi}$. 
As revealed by scrutinizing the RGEs in \eqref{eq:rge:ONscalarRGEs}, negative $\kappa_{\chi\xi}$ may keep the scalar part of portal coupling beta functions under control by generating negative contributions through mixed terms like in $\beta^{(1)}_{\kappa_{\phi\chi}} \supseteq 4N_\xi\kappa_{\phi\xi}\kappa_{\chi\xi}$.
Accordingly, for negative and sufficiently large $\kappa_{\chi\xi}$, there can be cancellations already within the scalar sector.
In addition, there exist combinations $(N_\chi,N_\xi)$, for which also the other free portal coupling, $\kappa_{\phi\chi}$ is of $\mathcal{O}(1)$. However, this is simply a consequence of the constraint derived from \eqref{eq:twoScalars:portalInequal}.

Finally, the most important result from the present paragraph is the following: the calculation based on the full set of RGEs with all terms included shows that none of the investigated models can be extrapolated all the way up to the Planck scale. Hence, we do not find a consistent conformal model in this class of theories.


\subsubsection{The minimal conformal model}
\label{sec:TwoScalars:MCM}
\noindent
As in the previous section, we will discuss the situation of two real scalar multiplets being added to the conformal SM. However, whereas earlier both additional scalars were supposed to have a trivial vacuum expectation value, we will now relax this assumption and investigate cases in which one of the multiplets has a component that acquires a \textit{finite} vev.

In the following, we will demonstrate that already for the simplest case with two additional scalar gauge singlets $S$ and $R$, the model allows for an extrapolation all the way up to the Planck scale, while giving rise to the correct phenomenology at the electroweak scale.
Since this time our goal is to actually prove that the largest possible UV scale is at least the Planck scale, it is no longer sufficient to calculate an upper bound for $\UVscale$ as we did before. In particular, we will not apply the previously introduced  \enquote{best-case approximation}.
Instead, we will use a two-step procedure: First, we determine the hypersurface in the model's parameter space on which the given low-energy phenomenology requirements are satisfied.
In particular and in contrast to our analyses before, we perform a fully consistent \textit{calculation} of the Gildener-Weinberg scale in the way outlined in Appendix \ref{app:calculatingGW}.
Second, we will numerically solve the \textit{full} set of RGEs towards the UV starting from the solution manifold from step one. 
At each RG step, we check if basic perturbativity and stability requirements are met by all running couplings.

Furthermore, we test whether no GW condition is satisfied at any intermediate energy scale $\Lambda>\GWscale$. During the evolution of the early universe, the tree-level potential would have developed a non-trivial minimum before reaching the original $\GWscale$ if such a scale $\Lambda$ existed. Hence, SSB would already have taken place at $\Lambda$ which would render our initial assumption inconsistent.

Let us now first concentrate on the conformal SM extended by two real scalar gauge singlets (CSM2S), one of which (say $S$) acquires a finite vev during EWSB, \ie $S = v_S + \sigma$. The most general scalar potential which is consistent with the SM gauge symmetries and classical scale invariance can be written as
\begin{align}
	\begin{split}
		V ={}& \lambda_\phi (\phi^\dagger\phi)^2 + \lambda_S S^4 + \lambda_R R^4 \\
		& + \kappa_{\phi S} (\phi^\dagger\phi) S^2 + \kappa_{\phi R} (\phi^\dagger\phi) R^2 + \kappa_{S R} S^2 R^2 \\
		& + \kappa_4 S R (\phi^\dagger \phi)^2 + \kappa_5 S^3 R + \kappa_6 S R^3 \fineq{.}
	\end{split}
	\label{eq:TwoScalars:potential-CSM2S}
\end{align}
In order to reduce the number of free parameters, we impose an additional global $\mathds{Z}_2$ symmetry in the following way
\begin{align}
	R \stackrel{\mathds{Z}_2}{\longrightarrow} - R \fineq{,}
	\label{eq:TwoScalars:Z2-CSM2S}
\end{align}
with \textit{all} other fields in the theory left invariant. The three terms in the last line of \eqref{eq:TwoScalars:potential-CSM2S} are odd under the above transformation and are thus forbidden. Note, furthermore, that the definition in \eqref{eq:TwoScalars:Z2-CSM2S} implies absolute stability of $R$ which therefore might be a viable dark matter candidate.\footnote{If we want the $\mathds{Z}_2$ symmetry to be exact, it \textit{must} be $R$ rather than $S$ which transforms non-trivially under it. Otherwise, $\mathds{Z}_2$ would be spontaneously broken by $v_S=\chevron{S} \neq 0$.}

As $R$ does not acquire a finite vev, it does not mix with the other CP-even scalar modes. With $\phi^0 = v_\phi + h/\sqrt{2}$ as usual, we then obtain the following tree-level mass
\begin{align*}
	m_R^2 = 2\left( \kappa_{\phi R} v_\phi^2 + \kappa_{SR} v_S^2 \right) \fineq{.}
\end{align*}
Furthermore, the $2\times 2$ mass matrix of $(h,\sigma)$ is the same as before in Section \ref{sec:oneScalar:real-wVEV}, \eqref{eq:oneScalar:massMatrix}, upon replacing
\begin{align*}
	\begin{split}
		\lambda_1 \to \lambda_\phi & \sepeq{,}
		\lambda_2 \to \lambda_S \fineq{,} \\
		\kappa_1 \to \kappa_{\phi S} & \sepeq{,}
		v_\chi \to v_S \fineq{.}
	\end{split}
\end{align*}
Using the above replacement rules, it is moreover straightforward to show that all formulas given in \eqref{eq:oneScalar:vevAlignment1} to \eqref{eq:oneScalar:angleRelationGW} apply to the present situation. In particular, we again have to distinguish positive and negative scalar mixing angle $\beta$ (\cf Table \ref{tab:oneScalar:twoCasesFiniteVEV}).

Here, we concentrate on the case in which the physical Higgs boson $\HiggsField$ is \textit{not} identified with the PGB of broken scale invariance. In this situation, the Higgs mass is given by $m_+$ from \eqref{eq:oneScalar:massEigenvaluesGW}. For given portal coupling $\kappa_{\phi S}(\GWscale)$, we can therefore directly calculate the value of $\lambda_\phi$ at the GW scale, namely
\begin{align*}
	\lambda_\phi(\kappa_{\phi S}) = \kappa_{\phi S} + \frac{\mHiggs^2}{4 v^2}  \quad \text{ at } \GWscale \fineq{.}
\end{align*}
We set $v_\phi \equiv v = \SI{174}{GeV}$ in accordance with the fact that $S$ is a gauge singlet whose vev does not contribute to the electroweak scale. The above equation can furthermore be used to determine the range of portal couplings consistent with positive $\beta$, namely $\absVal{\kappa_{\phi S}} < \num{0.065}$.

Next, we use the assumed vev configuration in form of the GW condition in \eqref{eq:oneScalar:GWconditions} to further reduce the number of free parameters at the initial scale:
\begin{align*}
	\lambda_S(\kappa_{\phi S}) = \frac{\kappa_{\phi S}^2}{4 \lambda_\phi} \quad \text{ at } \GWscale \fineq{.}
\end{align*}
The determination of the remaining parameters' initial values in terms of $\kappa_{\phi S}$ and $\mPGB$ is presented in Appendix \ref{app:calculatingGW}. In particular, we will show there how to consistently \textit{calculate} the GW scale.

Next, we need to clarify whether the model can be consistently extended all the way up to the Planck scale without any intermediate scale appearing. We do so by solving the theory's \textit{complete one-loop} RGEs. In each RG step, we check basic perturbativity and stability criteria of the model's couplings and abandon the given parameter point as soon as any inconsistency occurs below $\PlanckScale$.
The beta functions for the CSM2S can be obtained from the general formulas given in Appendix \ref{app:rge} by setting $N_\chi=N_\xi=1$.

\begin{figure}[t]
	\centering
	\includegraphics[width=0.97\columnwidth]{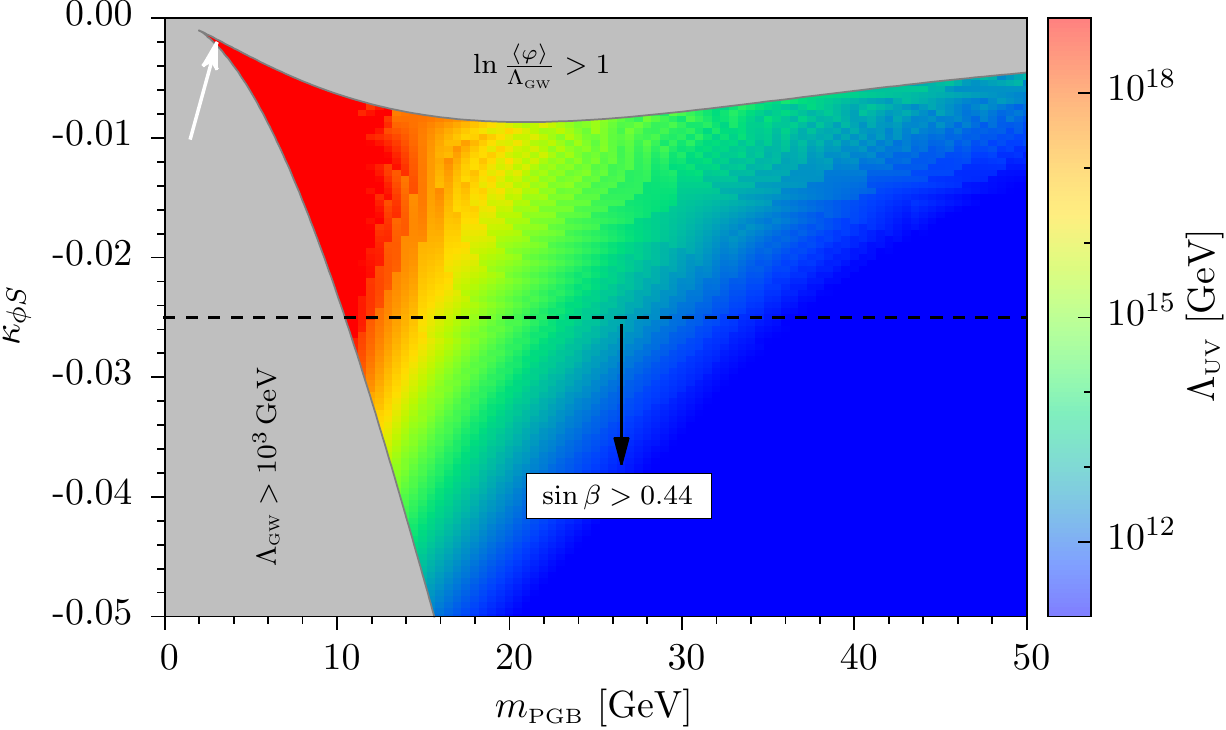}
	\caption{Largest possible UV scale in the CSM2S. (Case in which the physical Higgs is \textit{not} the PGB, \ie $\beta >0$.) The white arrow marks the example point from \eqref{eq:TwoScalars:MCMexamplePoint}.}
	\label{fig:TwoScalars:UVscale-CSM2S-HiggsNeqPGB}
\end{figure}

Our calculation for positive scalar mixing angle gives the plot shown in Figure~\ref{fig:TwoScalars:UVscale-CSM2S-HiggsNeqPGB}. As discussed above, we vary one of the portal couplings, $\kappa_{\phi S}$, and the PGB mass. In accordance with the discussion after \eqref{eq:oneScalar:portalExact}, we immediately discard those parameter points, which imply a large separation between $\GWscale$ and the electroweak scale $v$ (grey area on the left).
Since the effective potential's perturbative expansion is no longer reliable if $\ln ( \chevron{\varphi}/\GWscale)$ is too large, we additionally exclude points, for which the hierarchy between the GW scale and the condensate $\chevron{\varphi}$ becomes sizable (grey area on the top).
For small portal couplings $\absVal{\kappa_{\phi S}}$ and sufficiently low PGB masses, $\mPGB \lesssim \SI{15}{GeV}$, we then find a viable region of parameter space (red area). In this regime, a fully consistent extrapolation of the model up to the Planck scale is possible, while reproducing the correct low-energy phenomenology.

The available parameter space can be further narrowed down by noting that the mixing in the Higgs sector will effect the signal strength of Higgs events observed at the LHC. The currently measured signal strength constrains the scalar mixing angle to $\sin\beta \leq \num{0.44}$ \cite{Farzinnia:2013pga,Farzinnia2014}. By including this limit in Figure~\ref{fig:TwoScalars:UVscale-CSM2S-HiggsNeqPGB}, we can rule out all points below the dashed black curve.
Another type of constraint comes from the electroweak precision measurements performed at LEP.
However, as all new particles are scalar SM singlets, their contributions to the oblique $S$ parameter are necessarily both loop-suppressed and proportional to the small mixing angle $\beta$ \cite{Peskin1990}.
Corrections to the $T$ parameter are expected to be negligible as well since the model's scalar potential does not violate custodial symmetry.
A further interesting  phenomenological aspect is the existence of exotic Higgs decays. The Higgs boson can decay into two PGBs, which then further decay to SM particles. In this decay chain possible final states contain, $H\rightarrow 4 \text{ jets}$, $H\rightarrow 4 \text{ leptons}$, $H\rightarrow 4 \gamma$, $H\rightarrow 2 \text{ jets } 2 \gamma$, $H\rightarrow 2 \text{ jets } 2\text{ leptons}$, $H\rightarrow 2 \text{ leptons } 2 \gamma$. While the hadronic decays have a large background at the LHC, the final states containing leptons can be well distinguished. In particular the leptons are pairwise boosted in contrast to a decay mediated by the electroweak gauge bosons. Furthermore, the $H\rightarrow 4 \gamma$ can provide a very clean signature and only has the background coming from highly suppressed Higgs self interactions. This opens a window of opportunity to test a symmetry implemented close to the Planck scale, directly at the TeV scale.

\begin{figure}[t]
	\centering
	\includegraphics[width=0.97\columnwidth]{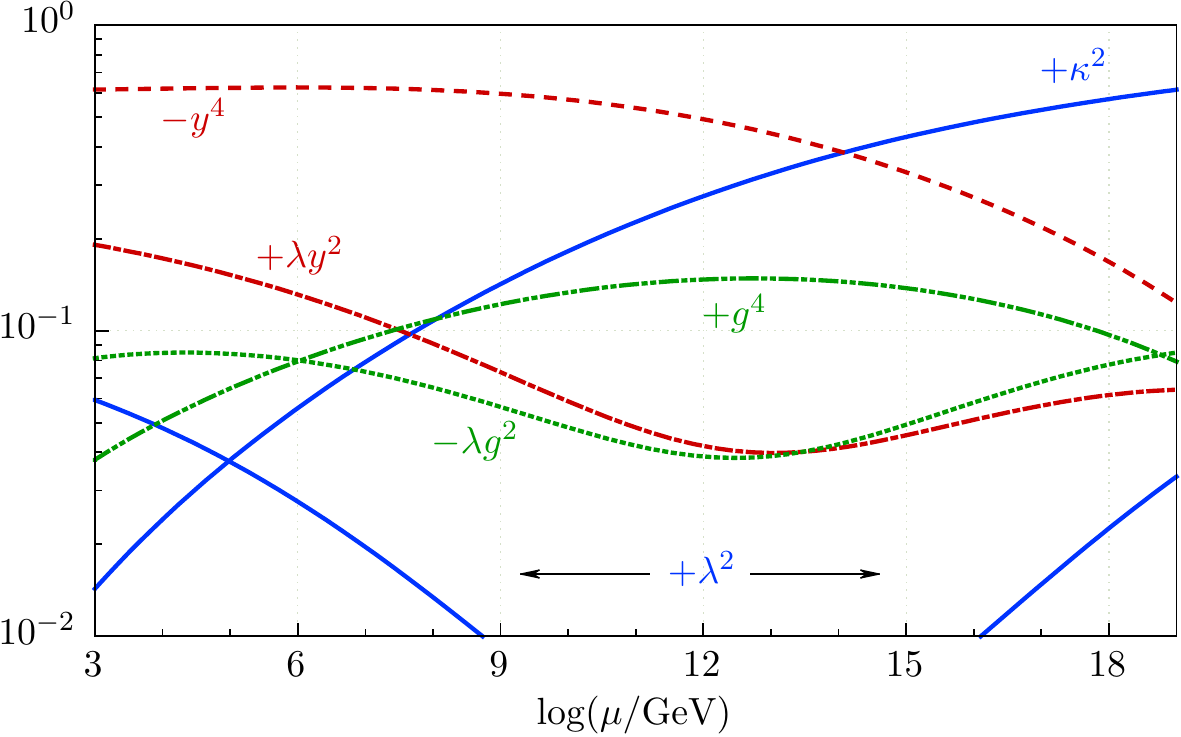}
	\caption{Running of the \textit{relative} contributions to the beta function of the Higgs self-coupling for the example point in \eqref{eq:TwoScalars:MCMexamplePoint}. The different contributions from the scalar, Yukawa and gauge sectors are displayed in blue, red and green, respectively (\cf also caption of Figure~\ref{fig:oneScalar:betaFctContrib}).}
	\label{fig:TwoScalars:betaContrib-CSM2S-HiggsNeqPGB}
\end{figure}

Let us now try to gain further insight on how the scalar couplings can remain free of Landau poles in the present model. In analogy to our analysis in Section~\ref{sec:oneScalar:real-woVEV}, we therefore compare the different contributions to the Higgs self-coupling beta function.
The corresponding results presented in Figure~\ref{fig:TwoScalars:betaContrib-CSM2S-HiggsNeqPGB} were obtained for the example point marked in Figure~\ref{fig:TwoScalars:UVscale-CSM2S-HiggsNeqPGB}, namely
\begin{align}
	\mPGB = \SI{3}{GeV} \sepeq{,}
	\kappa_{\phi S} = \num{-0.0018} \fineq{.}
	\label{eq:TwoScalars:MCMexamplePoint}
\end{align}
Requiring the correct vev implies $m_R = \SI{313}{GeV}$. Stable RG running up to the Planck scale is then \eg possible for $\lambda_{R} = \num{0.015}$ and $\kappa_{S R} = \num{0.01}$.

Now, the key difference with respect to Figure~\ref{fig:oneScalar:betaFctContrib} is that the pure scalar contribution no longer dominates over the whole energy range. Rather, it is exceeded by the stabilizing contribution from the Yukawa coupling for energies up to $10^{14}\,$GeV so that the Higgs coupling first decreases.
At larger scales, the portal terms start to dominate. The coupling will hence ultimately develop a Landau pole. However, our calculation shows that $\lambda_\phi$ stays small up to the Planck scale.

We attribute this improved behavior of the scalar contribution to two features.
First, compared to the models in Section~\ref{sec:oneScalar}, there is now a larger number of \textit{independent} scalar degrees of freedom (multiplets). Positivity of $B$ therefore no longer implies that one coupling must be particularly large at the initial scale: Whereas \eg equations (\ref{eq:oneScalar:portalExact}) and (\ref{eq:oneScalar:allowedRange-real-wVEV}) require $\kappa_1$ to be of order one, the corresponding portal coupling in the CSM2S, $\kappa_{\phi S}$, is preferably of $\mathcal{O}(10^{-2})$ or smaller.
Second, since all scalar couplings can now be of the same order of magnitude, there exists the possibility of cancellations between different scalar contributions. Those cancellations may help to keep the beta functions of the portal couplings small.

\begin{figure}[t]
	\centering
	\includegraphics[width=0.97\columnwidth]{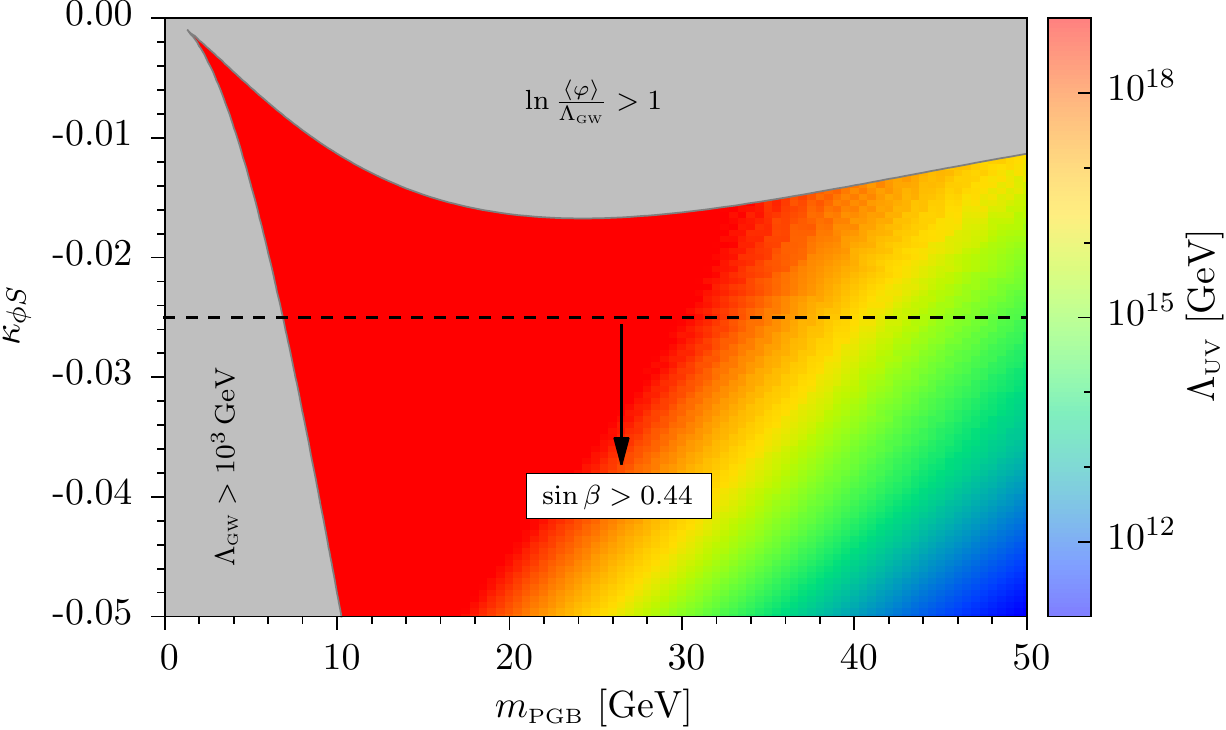}
	\caption{Largest possible UV scale in the CSMTS. (Case in which the physical Higgs is \textit{not} the PGB, \ie $\beta >0$)}
	\label{fig:TwoScalars:UVscale-CTSMS-HiggsNeqPGB}
\end{figure}

Now that we have understood how a stable RG running is realized in the minimal conformal model, let us look for means to achieve larger PGB masses. To that end, we briefly discuss a non-minimal extension of the conformal SM very similar to the CSM2S. Here, the singlet scalar $R$ is exchanged for a real SU(2)$_L$ triplet with vanishing vev.\footnote{The imposed $\mathds{Z}_2$ symmetry from \eqref{eq:TwoScalars:Z2-CSM2S} is replaced by a global $O(3)$ symmetry in the triplet sector.} The resulting model will be referred to as CSMTS in the following.
Figure~\ref{fig:TwoScalars:UVscale-CTSMS-HiggsNeqPGB} demonstrates that compared to the minimal conformal model an extended region of PGB masses up to $\mPGB \approx \SI{35}{GeV}$ becomes accessible in the CSMTS. With respect to the minimal model, two heavy scalar degrees of freedom are added to the theory's spectrum.
According to \eqref{GW_formalism_PGBmass} and \eqref{GW_formalism_LoopFunctionsAB}, a given PGB mass can now be produced for smaller initial values of the scalar couplings (\cf Appendix~\ref{sec:GWformalism}). Consequently, potential Landau poles will develop at higher scales. A straightforward and minimally invasive way to generate even larger PGB masses would be to replace the triplet by a higher-dimensional real SU(2)$_L$ multiplet, \eg a septet.
Note that in this case the dark matter stability does not need to be enforced by any additional global symmetry.

Let us finally comment on the robustness of our results under inclusion of higher loop orders in the RG running. Since higher-order terms come with an additional loop suppression factor of $1/16\pi^2$, their contributions can only have a significant impact, if the one-loop beta functions are anomalously small.
Hence, in all the cases that failed to provide a perturbative evolution up to $\PlanckScale$, two-loop effects will be negligible, since the one-loop beta functions are already sizable.
In contrast, as there are mild cancellations in the RGEs of the minimal conformal model, two-loop contributions may change our results quantitatively. 
If two-loop contributions turned out to be sizeable in some areas of parameter space with mild cancellations between the one-loop contributions, the former might destabilize the RGE running. In such cases the affected parameter space would need to be excluded.
However, we expect our findings to remain valid from a qualitative perspective.

\section{Matching to the Semiclassical Regime in Gravity}
\label{sec:SemiclassicalGravity}
\noindent
In this section we sketch how our scenario might be embedded in a broader context including effects from gravity. We choose here the semi-classical approach to quantum fields in curved space-time as it is the most conservative method currently available and only requires concepts of general relativity and quantum field theory. 
We have seen in the previous section that small or even vanishing self interactions in the scalar sector lead to stable RG trajectories. Motivated by this observation we consider a free scalar field in the gravitational background. As a model system we consider de-Sitter geometry as we assume that it is a good description for the early state of our universe. The metric ansatz is conveniently parametrized by $\text{d}s^2 = a(t)^2 (\text{d}\eta^2 - \text{d}\vec{x}^2)$ where $\text{d} \eta = \text{d}t/a(t)$ is the conformal time coordinate. The governing equation in this highly symmetric system is the trace of the Einstein equation, given by 
\begin{align}
\label{eqn:EinsteinTrace}
R \, \frac{\PlanckScale^2}{8 \pi} = - \langle T_{ \,\, \mu}^\mu \rangle\,.
\end{align} 
where $R$ denotes the Ricci curvature scalar, which represents the gravitational field.  The vacuum expectation value of the scalar-field EMT sources the gravitational field and is given in four space-time dimensions by
\begin{align}
T_{ \,\, \mu}^\mu  = \frac{1}{2} m^2 \phi^2  + \frac{3}{2} \left( \xi - \frac{1}{6}\right)\Box \left( \phi^2 \right) \,. 
\end{align} 
As we assume conformal initial conditions with $m=0$ and the coupling of the scalar $\phi$ to the curvature $\xi = 1/6$, this quantity is zero at tree level and we need to compute its vacuum expectation value quantizing the scalar field $\phi$. We will only sketch the slightly technical calculation at this point and make reference to the literature for a more pedagogical description \cite{Parker:1978gh,Smirnov:2014}. The basic idea is that we construct the field operators as 
\begin{align}
\phi(x) = \sum_{\vec{k}} \left( A_{\vec{k}} f_{\vec{k}}(x) + A^\dagger_{\vec{k}} f^{*}_{\vec{k}}(x) \right)\,.
\end{align}
In the above equation the mode functions $f_{\vec{k}}(x)$ are the solutions to the equation of motion for the scalar field in curved background. The ladder operators $A_{\vec{k}}$, $A_{\vec{k}}^\dag$ define the vacuum by $A_{\vec{k}} |0\rangle = 0$ for all $\vec{k}$. This vacuum state is called the adiabatic vacuum as it is assumed that the components of the metric tensor change in such a way that we can define a sensible expansion in the components' derivatives. In our case of de-Sitter geometry this translates into an assumption about the scale factor and its time derivatives. We begin with a general form of the solution for the sclar field equation of motion
\begin{align}
f_k = \frac{1}{\sqrt{2\,V}} \, g(a)\, h_k \,e^{i\,\vec{k}\cdot \vec{x}}.
\end{align}
The rescaling function $g(a)$ will drop out of the vacuum expectation values of operators quadratic in fields and thus has no physical meaning. This includes the vacuum expectation value of the EMT, which we will compute below. At the same time $g(a)$ can be chosen such that the friction term in the general equation of motion is not present and transforms it into a harmonic oscillator equation with time-varying mass. So choosing conformal time coordinate $\eta$, the rescaling needs to be $g(a) = 1/a(\eta)$ in order to cancel out the friction term. We find that with this parametrization the equation for $h(\eta)$ is as follows 
\begin{align}
\label{eqn:MassTimeDependent}
h_k'' + \Omega_k^2 h_k = 0 \, ,
\end{align}
where the oscillation frequency is defined as
\begin{align}
\begin{split}
\Omega_k^2 &=  k^2 + \left( m^2 + \left(\xi -\frac{1}{6}\right)R\right)\,a(\eta)^2 
\\
&\equiv k^2 + m_\text{eff}^2 \, a(\eta)^2 \, .
\end{split}
\end{align}
In the above equation, $R$ is the Ricci scalar of the Friedmann-Robertson-Walker (FRW) spacetime and we have introduced the effective mass parameter $m_\text{eff}$.
This parametrization shows immediately the special case of the de-Sitter spacetime: if $R$ is constant the curvature-induced term amounts to a mass correction.

The solution to the equation of motion for each mode $f_{\vec{k}}$ can be found in an adiabatic series. When substituted in the equation for the EMT between two vacuum states and summing over all modes, it leads after renormalization to an expression in the de-Sitter background%
\footnote{The general form of the trace of the energy-momentum tensor can for example be found in \cite{Anderson:1987yt}. Their Eq.~(3.2) reduces to our expression assuming the de-Sitter symmetries.}
\begin{align}
\label{eqn:TracedeSitter}
	\begin{split}
		\bra{0} T^\mu_{\,\,\,\,\mu} \ket{0}= \frac{1}{16\pi^2}\left( -\frac{1}{2160} \,R^2 + \frac{m^2}{18} \,R  \right. \\
		\left. +\frac{1}{3} \left(\xi -1/6\right)m^2 \,R + \frac{1}{2}\left(\xi -1/6\right)^2 \,R^2\right)\,.
	\end{split}
\end{align} 
As discussed, our initial conditions were chosen to be $m=0$ and $\xi =1/6$. Thus $16 \, \pi^2 \bra{0} T^\mu_{\,\,\,\,\mu} \ket{0}= -\frac{1}{2160} \,R^2$, which is called the gravitational conformal anomaly.%
\footnote{Note that we provide here a toy example of a scalar coupled to gravity in order to demonstrate our proposed mechanism. In a full SM context the contribution to the EMT due to the non-vanishing beta function of the hypercharge \textendash\ proportional to $\chevron{F_{\mu \nu} F^{\mu \nu}}$ \textendash\ is present and would have to be canceled by the gravitational conformal anomaly as well.}

From \eqref{eqn:EinsteinTrace} it is clear that this vacuum set-up leads to an inflationary solution with a constant space-time curvature and a scale factor time evolution $a(t) \propto \exp{ \left(H\,t \right)}$, where $H = \sqrt{R}/12$ is the Hubble rate. At first glance this might seem as inflation would continue forever, but as the space-time expands the temperature drops, which changes the energy scale and induces a running of the parameters. As we discussed in the introduction, even at vanishing quartic interaction of the scalar the running of the gauge couplings translates into a running of $\xi$ at higher orders, leading to a deviation of $\xi$ from the value $1/6$. 

Since the contribution of the non-minimal coupling $\xi$ to the trace of the EMT is positive definite [\cf \eqref{eqn:TracedeSitter}], it unavoidably cancels the contribution of the gravitational conformal anomaly. This ends the inflationary epoch and allows therefore the universe to enter its later FRW evolution. It is important to evaluate the scale evolution of the effective mass parameter of the scalar field during this process. As mentioned, at the beginning $m=0$ and $\xi= 1/6$ which means that $m_\text{eff} = 0$. Then, once $\xi$ deviates by a value $\epsilon$ from $1/6$ we can use \eqref{eqn:EinsteinTrace} to infer that  $m_\text{eff} \approx 2 M_\text{pl} \sqrt{\pi \epsilon \left(1080^{-1} - \epsilon^2\right)^{-1}}$. This is a valid approximation for a non-vanishing trace of the EMT. At the same time in the limit $ \langle T_{ \,\, \mu}^\mu \rangle \rightarrow 0$ \eqref{eqn:EinsteinTrace} shows that $R \rightarrow 0$ and thus $m_\text{eff} \rightarrow 0 $. This point in the evolution is special, as the trace of the EMT vanishes even at quantum level. Therefore, at the end of inflation a transition to the FRW radiation-dominated epoch takes place and $m_\text{eff} \approx 0 $ with corrections of order $ \epsilon H_\text{reheating}$. Under the reasonable assumption that at reheating  the space-time curvature scale $H_\text{reheating}$ is much smaller than the electroweak scale the boundary condition of vanishing explicit masses is a good approximation for our study of the electroweak sector and scale invariance turns out to be an approximative symmetry with corrections of order  $ \epsilon H_\text{reheating}$.

We argue that this cosmological scenario is a good motivation for our field theory set-up with a classically vanishing mass and asymptotically small quartic self interactions of the scalar fields.  
Note that we did not rely on a loop expansion to arrive at this conclusion, but rather used the adiabatic expansion in metric derivatives.


\section{Discussion}
\label{sec:implications}
\noindent
The present study contains the analysis of simple conformal extensions of the Higgs sector in which radiative symmetry breaking within the Coleman-Weinberg mechanism can take place. As a consequence of non-linearly realized conformal symmetry implemented at a much higher scale, the usual gauge hierarchy problem is avoided. For this scenario to be consistent, the vanishing of the trace anomaly at the high scale is necessary. We discuss how this scenario can be realized by a semiclassical matching to gravity in Section \ref{sec:SemiclassicalGravity}.

As simple extensions of the Standard Model (SM), we consider theories with the same gauge group. Hence, there is always the beta function of the Abelian gauge coupling which can only vanish in the UV once gravity contributions become significant. Thus, our necessary condition is that the renormalization-group (RG) running remains stable and does not develop Landau poles below the Planck scale. We have used the Gildener-Weinberg formalism, ensuring the perturbative nature of our expansion, and have taken into account the complete one-loop RG equations. In particular, we include contributions from field renormalization.

We find that none of the conformal extensions of the Higgs sector by one scalar SU(2)$_L$ multiplet meets the stability criteria. The additional scalar can be either real or complex and acquire a vacuum expectation value or not. In all cases the models develop a Landau pole far below the Planck scale. The reason is that in all parameter points the phenomenological requirement that the Higgs boson mass is roughly half its vacuum expectation value, leads to large portal couplings $\kappa$ in the potential. The RG running is then highly unstable, since the beta function for the Higgs quartic coupling $\lambda$ contains terms proportional to $\lambda^2$ and $\kappa^2$ with positive coefficients. Contributions from gauge bosons can decelerate the running, as they contain negative terms of the form $- \lambda g^2$, where $g$ is a generic gauge coupling. However, with growing scalar couplings the scalar sector dominates and the system is still unstable. 

In particular, the simple model discussed in \cite{Meissner:2006zh}, in  which the SM is extended by one real SU(2)$_L$-singlet scalar and right-handed sterile neutrinos, turns out to be unstable. Indeed, even though a Yukawa coupling $y$ gives a negative contribution proportional to $-y^4$ to the beta function of the Higgs self-coupling, the scalar field wave function renormalization unavoidably introduces positive terms scaling as $+\lambda y^2$. Therefore, it is obvious that, with growing $\lambda$, the fermionic contributions always destabilize the system even more.\footnote{The different treatment of the wave function renormalization leads to deviations with respect to the results of \cite{Meissner:2006zh}.}

Other extensions of the Higgs sector by one SU(2)$_L$ scalar representation turn out to be unstable as well, as for example the conformal inert doublet model \cite{Hambye:2007vf}.\footnote{In order to check this, we have performed a fully consistent analysis as presented in Section~\ref{sec:TwoScalars:MCM}. In particular, we have taken into account the complete scalar potential including the term 
$\Delta V_2  = \kappa_4 \bigl[ (\phi^\dagger \chi)^2 + \hc \bigr]$, which is only present for $\chi \sim (1, \,2, \, \tfrac{1}{2})$.}

Having excluded those simplest theories, we find the minimal model, which leads to correct radiative breaking of electroweak symmetry and is RG stable, among the extensions of the Higgs sector by two scalars. To be precise, our analysis shows that the minimal model is the SM augmented by two scalar gauge singlets, one of which has to obtain a non-zero vacuum expectation value. In this system a light Higgs boson can be realized without fine-tuning. In addition, the theory contains a pseudo-Goldstone boson (PGB) with its mass being strongly suppressed with respect to the vacuum expectation value of the new singlet scalar. This turns out to be a natural set-up with no need for large couplings in the potential. Furthermore, in the three-scalar potential the portal term contributions to the RG running can be negative and thus mutually stabilize their beta functions. Those are the two reasons why the system remains stable up to the Planck scale. 

Our study raises the general question about the stability of a Standard Model extension under RG translations. A general observation is that the top-quark Yukawa coupling runs towards a stable value between $0.4$ and $0.6$ in the far ultraviolet, depending indirectly on the SU(2)$_L$ scalar content.
This is due to the fact that the top Yukawa beta function at one-loop depends on itself and the three gauge couplings, which only show a mild running in the ultraviolet regime. As can be seen from the appropriate RG equations, the Higgs quartic coupling can have a regime of RG-flow stability at finite values given a large top Yukawa and small portal couplings to the new scalars.
It is an interesting and non-trivial observation that in the SM there are no Landau poles below the Planck scale and the Higgs self-coupling approaches a constant (yet negative) value in the UV. In our extensions of the Higgs sector portal couplings are necessarily present. Their positive contributions to the Higgs beta function lead to vacuum stability at all energy scales.
Additionally, if the portal couplings are sufficiently small, the near vanishing of the Higgs beta function in the UV is maintained.\footnote{The exact value of the Higgs quartic coupling at which its beta function vanishes is sensitive to the top quark mass and can have a small positive value or even vanish for some top mass values \cite{Holthausen:2011aa}.}
Accordingly, in the RG-stable region of parameter space the scalar beta functions have very small values at the Planck scale. This is the desired behavior to achieve the necessary anomaly matching, thus indicating that conformal symmetry is realized at the quantum level. 

Within the minimal model, we find that one of the scalar singlets is an excellent dark matter candidate, since it does not develop a vacuum expectation value. Its effective phenomenology is similar to the Higgs portal model, see for example \cite{Duerr:2015bea,Duerr:2015aka,Duerr:2015mva} and references therein. We observe the dark matter mass to be confined to a rather small region between \SI{300}{GeV} and \SI{370}{GeV}. Furthermore, we checked that the parameter space considered by us is consistent with cosmological observations, \ie the scalar field abundance does not overclose the universe. However, a detailed study of the dark matter phenomenology goes beyond the scope of this article. We stress again that the stability of the DM candidate crucially relies on the assumed $\mathds{Z}_2$ symmetry. In contrast, if the second scalar is a septet its stability does not need to be enforced by any additional symmetry.

Another important phenomenological consequence is that the mass of the pseudo-Goldstone boson is found to be always below half the Higgs mass and is preferably as light as a few GeV. This necessarily leads to additional Higgs decays and therefore to a larger Higgs width than in the SM. 

Furthermore, the points of the parameter space in which we find stable RG running predict sizable singlet scalar admixtures to the physical Higgs state with sines of the mixing angle between $0.12$ and $0.48$.
The mixing can be compared to the SM prediction which leads to a constraint on the mixing angle. The current LHC upper limit of $\sin \beta \leq \num{0.44}$ \cite{Farzinnia:2013pga,Farzinnia2014} therefore already rules out a certain fraction of the parameter space. The complete model might be tested by the LHC in the ongoing run.

Finally, we would like to remark that in the minimal conformal extension of the SM neutrino masses can easily be accommodated. Once we introduce right-handed neutrino fields as SM gauge singlets they naturally possess a conformal and gauge-invariant, Majorana-type Yukawa coupling to the scalar singlet $S$.\footnote{Note that the right-handed neutrinos, however, do not couple to $R$ due to the $Z_2$ symmetry. Even if such a coupling existed, it still would not lead to a Majorana mass term because $R$ does not develop a finite vev.}  Additionally, we obtain Dirac-type Yukawa couplings with the SM lepton and Higgs doublet. After electroweak symmetry breaking the Yukawa couplings lead to a neutrino mass matrix that realizes a type-I seesaw mechanism \cite{Lindner:2014oea}. 
Of course, it remains to be checked whether including the Majorana Yukawa coupling negatively influences the RG running. Based on our observations regarding the effects of the top quark Yukawa coupling on the RGEs, we expect changes due to $y_\tinytext{M}$ to be controllable. 

To summarize our results, we found that it is necessary to add at least two scalar fields to the Standard Model, one of which has to develop a non-vanishing vacuum expectation value to have a model which is stable under RG translations. Thus the minimal model we discuss is an extension of the SM Higgs sector by two real singlet scalar fields. We have found that the minimal model contains a viable dark matter candidate and predicts sizable mixing in the Higgs sector, which might be a powerful tool to rule out or get a hint about the realization of conformal models.

\section*{acknowledgements}
\noindent
We acknowledge very helpful discussions with Hermann Nicolai and his comments on our manuscript.
We thank Pavel Fileviez Perez for discussions.

\appendix


\section{The Gildener-Weinberg formalism}
\label{sec:GWformalism}
\noindent
In this appendix, we review the Gildener-Weinberg formalism introduced in \cite{Gildener1976b}. 
Within the framework of the Coleman-Weinberg mechanism, it allows to systematically minimize a potential in a theory with multiple scalar fields without having to resort to numerical brute-force algorithms.
It can be considered the analogue of the RG-improved potential in the one-scalar case.
Furthermore, the formalism allows us to derive conditions on the scalar couplings and ensures the applicability of the loop expansion of the effective potential.

After discussing the formalism's basic principles in Section~\ref{app:basics}, we present our method to consistently calculate the scale of spontaneous symmetry breaking $\GWscale$ in Section~\ref{app:calculatingGW}.

\subsection{Basics}
\label{app:basics}
\noindent
Gildener and Weinberg start by introducing a general tree-level scalar potential of the form 
\begin{align}
	V (\myVec{\Phi})
	=
	\frac{1}{24} \, f_{i j k \ell} \, \phi_i \, \phi_j \, \phi_k \, \phi_\ell
	\, ,
	\label{GW_formalism_GeneralPotial}
\end{align}
where $\myVec{\Phi}$ denotes the collection of all real scalar degrees of freedom in a given theory. Note that the coupling constants $f = f (\Lambda)$ are subject to RG running, where $\Lambda$ is the renormalization scale. The potential \eqref{GW_formalism_GeneralPotial} is assumed to develop a continuous set of degenerate minima at a specific scale $\GWscale$, the Gildener-Weinberg scale. These minima lie on a ray through the origin of field space, henceforth referred to as the \textit{flat direction}.

In order for $V$ to develop a flat direction, the scalar couplings must satisfy certain conditions which can generically be written as
\begin{equation}
	\Bigl.
		R (f) \,
	\Bigr|_{\Lambda = \GWscale}
	=
	0
	\, ,
	\label{GW_formalism_MinimumCondition}
\end{equation}
thereby determining the GW scale. We will refer to conditions of this type as Gildener-Weinberg conditions.
The flat direction can be parametrized as
\begin{equation}
	\myVec{\Phi}_\text{flat}
	=
	\myVec{n} \varphi
	\, ,
	\label{GW_formalism_FlatDirection_Definition}
\end{equation}
where $\myVec{n}$ is a unit vector and $\varphi$ gives the position on the ray.
Whereas the tree-level potential is minimal for each $\varphi$, loop corrections will in general bend the potential along the flat direction. Thus, a particular value $\chevron{\varphi}$ is singled out as the actual minimum.
Equation (\ref{GW_formalism_FlatDirection_Definition}) then implies that the scalar fields corresponding to the non-vanishing components of $\myVec{n}$ acquire finite vevs, the relative magnitudes of which are given by the entries of $\myVec{n}$.
Depending on which of the scalar modes acquire a finite vev, the relevant set of conditions $R$ is different.

The one-loop effective potential along the flat direction can be written as \citep{Gildener1976b}
\begin{align}
	V_\text{eff}^{(1)}(\myVec{n} \varphi) 
	= 
	A\varphi^4 + B\varphi^4 \ln \left( \frac{\varphi^2}{\GWscale^2} \right)
	\, ,
	\label{GW_formalism_EffectivePotential_1Loop}
\end{align}
where $\GWscale$ is the renormalization point.%
\footnote{Notice that due to dimensional transmutation all dimensional quantities, and in particular masses, will be proportional to the symmetry breaking scale. Hence, it is only reasonable to take $\GWscale$ as the renormalization point in \eqref{GW_formalism_EffectivePotential_1Loop}.}
The functions $A$ and $B$ are given by
\begin{widetext}
\begin{align}
	A 
	= 
	\frac{1}{64 \pi^2 \chevron{\varphi}^4} 
	\sum_i  
	(-1)^{2s_i} d_i \cdot m_i^4 \, (\myVec{n} \chevron{\varphi}) 
	\left( 
		\ln \frac{m_i^2(\myVec{n}\chevron{\varphi})}{\chevron{\varphi}^2} 
		- c_i 
	\right) 
	\, ,
	&&
	B 
	= 
	\frac{1}{64 \pi^2 \chevron{\varphi}^4} 
	\sum_i 
	(-1)^{2s_i} d_i \cdot m_i^4 \, (\myVec{n} \chevron{\varphi}) 
	\, .
	\label{GW_formalism_LoopFunctionsAB}
\end{align}
\end{widetext}
A few comments on the notation are in order. First, the index $i$ in the above sums runs over all particles in the given theory. For each particle $m_i (\myVec{n}\varphi)$ is given by its field-dependent \textit{tree-level} mass evaluated along the flat direction. Note that $m_i$ implicitly depends on the renormalization point $\GWscale$.
The coefficient $d_i$ counts the particle's real degrees of freedom and $s_i$ denotes its spin. The constants $c_i$ depend on the actual renormalization scheme. Here, we will use the $\overline{\text{MS}}$ scheme, for which one finds $c_i=\tfrac{5}{6}$ in the case of gauge bosons and $c_i=\tfrac{3}{2}$ for scalars or fermions.
Finally, as mentioned before, $\chevron{\varphi}$ is the value of the parameter $\varphi$ along the flat direction at which the one-loop effective potential develops an extremum. This extremum is a minimum if and only if $B$ is positive.
In particular, it is straightforward to show that the minimum of the one-loop effective potential \eqref{GW_formalism_EffectivePotential_1Loop} along the flat direction lies at
\begin{equation}
	\chevron{\varphi} 
	= 
	\GWscale \cdot \exp \left( -\frac{1}{4} - \frac{A}{2B} \right)
	\, .
	\label{GW_formalism_condensate}
\end{equation}
The above equation shows that $\chevron{\varphi}$ is of the same order as $\GWscale$ if $A$ is of the same order as $B$. This is a necessary condition to control the loop expansion in powers of $\ln ( \chevron{\varphi} / \GWscale )$.

The excitation along the flat direction $\myVec{\Phi}_\text{flat}$ defined in \eqref{GW_formalism_FlatDirection_Definition} is the pseudo-Goldstone boson of broken scale invariance. Massless at tree level, its mass is generated radiatively only after SSB. The mass of the PGB at one-loop level is given by
\begin{align}
	m_\tinytext{PGB}^2 
	= 
	\left. 
		\frac{\dd^2 V^{(1)}_\text{eff}(\myVec{n}\varphi)}{\dd \varphi^2}
	\right|_{\varphi = \chevron{\varphi}} 
	= 
	8 B \chevron{\varphi}^2 
	\, .
	\label{GW_formalism_PGBmass}
\end{align}
Note that in models, in which the PGB is identified with the Higgs boson measured at the LHC, $\HiggsField$, the loop function $B$ has to match the Higgs mass according to this equation.

\subsection{Calculating the Gildener-Weinberg scale}
\label{app:calculatingGW}
\noindent
In this part, we enlarge upon certain aspects of the Gildener-Weinberg formalism introduced in the previous section. Thereby, we concentrate on the consistent computation of the GW scale, which we need in our treatment of the minimal conformal model in Section~\ref{sec:TwoScalars:MCM}.
There, we have already described how to express $\lambda_\phi(\GWscale)$ and $\lambda_S(\GWscale)$ in terms of $\kappa_{\phi S}(\GWscale)$. Now, we show how to calculate $\GWscale$ and $m_R$ in a way consistent with the empirically known values of $v$ and $\mHiggs$ for given $\kappa_{\phi S}$ and $\mPGB$.

The crucial quantity in determining a viable parameter point is the loop function $B$ introduced in \eqref{GW_formalism_LoopFunctionsAB}. It is particularly important since it relates the PGB mass to the other particles' masses and the condensation scale via \eqref{GW_formalism_PGBmass}. As a first step, we isolate the contributions due to SM fermions and gauge bosons,
\begin{align*}
	B = \BSM + B_\text{add} \fineq{,}
\end{align*}
where $B_\text{add}$ contains all additional contributions from the scalar sector, including the one from the SM Higgs doublet.

Before we proceed let us remark that the models we consider in Section~\ref{sec:TwoScalars:MCM} are special in the following sense: In addition to the usual Higgs doublet only gauge singlets obtain non-vanishing vevs. Hence the electroweak scale originates from the doublet sector only, $v = v_\phi = n_1 \chevron{\varphi}$, and we can therefore parametrize all SM fermion and gauge boson masses as
\begin{equation}
	m_i
	=
	\tilde{m}_i \, n_1 \chevron{\varphi}
	\text{~~~for~~}
	i \, \in \, \text{SM}^*
	\label{Appendix_MassTermRunningSM}
\end{equation}
with appropriate dimensionless coefficients $\tilde{m}_i$. The set SM$^*$ contains all massive SM gauge bosons and fermions.
Defining the function
\begin{align*}
	\tilde{B}_\tinytext{SM}
	=
	\frac{1}{64 \pi^2} 
	\sum_{i \in \tinytext{SM}^*} 
	(-1)^{2s_i} d_i \cdot \tilde{m}_i^4
	\, ,
\end{align*}
and using \eqref{Appendix_MassTermRunningSM}, we can write
\begin{align}
	B(\GWscale) = n_1^4 \BSMtilde(\GWscale) + B_\text{add} \fineq{.}
	\label{eq:calculatingGW:B}
\end{align}
The above partition is particularly convenient, because $n_1$ and $B_\text{add}$ only depend on the scalar couplings whose values are \textit{defined at the GW scale}. We suppress this implicit dependence on $\GWscale$ in the above equation.
In contrast, $\BSMtilde$ depends on SM gauge and Yukawa couplings, which are only \textit{known at the electroweak scale}. However, we can use the RGEs to evolve the gauge and Yukawa couplings to any scale $\Lambda < \PlanckScale$.

Motivated by combining \eqref{eq:calculatingGW:B} with the formula for the PGB mass, \eqref{GW_formalism_PGBmass}, we define the function
\begin{align}
	\begin{split}
		G(\Lambda) := n_1^4 \BSMtilde(\Lambda) 
		+ \underbrace{B^\prime_\text{add} + \frac{n_1^4}{64 \pi^2} \frac{\mHiggs^4}{v^4}}_{= B_\text{add}} - \frac{n_1^2}{8} \frac{\mPGB^2}{v^2} \fineq{.}
	\end{split}
	\label{eq:calculatingGW:funcG}
\end{align}
Then the Gildener-Weinberg scale consistent with a given set of scalar couplings and a particular PGB mass is defined via the condition\footnote{The uniqueness of the root of \eqref{eq:calculatingGW:consistency1} is guaranteed, since $\BSMtilde(\Lambda)$ is a strictly monotonously increasing function of the energy scale $\Lambda$.}
\begin{align}
	G(\GWscale) \stackrel{!}{=} 0 \fineq{.}
	\label{eq:calculatingGW:consistency1}
\end{align}
In addition, we must check whether $\GWscale$ and $v$ are reasonably close to each other (\cf our discussion after \eqref{eq:oneScalar:portalExact}).

In our previous discussion, we have already assumed the electroweak scale to attain its proper value, $v=\SI{174}{GeV}$. For a fully consistent calculation, we must therefore ascertain, if our one-loop effective potential, indeed, possesses an appropriate minimum, \ie whether  
\begin{align}
	v =
	n_1 \GWscale \cdot \exp \left( -\frac{1}{4} - \frac{A}{2B} \right)
	\fineq{}
	\label{eq:calculatingGW:consistency2}
\end{align}
yields the correct number (\cf \eqref{GW_formalism_condensate}).

Solving \eqref{eq:calculatingGW:consistency1} and \eqref{eq:calculatingGW:consistency2} self-consistently, eventually fixes two more variables at the GW scale, namely $\GWscale$ itself and $B^\prime_\text{add}$. Of course, the particular form of $B^\prime_\text{add}$ depends on the model under investigation. An explicit calculation in the minimal conformal model from Section~\ref{sec:TwoScalars:MCM}, for instance, gives
\begin{align*}
	B^\prime_\text{add} = \frac{\kappa_{\phi R}n_1^2 + \kappa_{S R}n_2^2}{16 \pi^2} \fineq{.}
\end{align*}


\section{One-loop beta functions} \label{app:rge}
\noindent
In this appendix, we collect the renormalization group equations used throughout this work. For the perturbative expansion of a given beta function, we adopt the following notation
\begin{align*}
	\beta(g) = \frac{\beta^{(1)}(g)}{16\pi^2} + \frac{\beta^{(2)}(g)}{(16\pi^2)^2} + \ldots \fineq{.}
\end{align*}
In the following we list the one-loop beta functions for two real scalars (\ref{app:rge:SM+real}) and one complex scalar (\ref{app:rge:SM+cplx}). The beta functions for one real scalar are obtained from the two-scalar functions by dropping every term with a $\xi$.


\subsection{SM + real scalar representation(s)} \label{app:rge:SM+real}
\noindent
Here, we consider the scalar sector of the SM supplemented by up to two \textit{real} scalar SU(2)$_L$ multiplets, denoted as $\chi$ and $\xi$, respectively.
For the following, recall the definition of the quadratic Casimir $C$ and the Dynkin index $D$, which are, respectively, given by
\begin{align}
	C = \tfrac{1}{4} (N^2-1) \sepeq{,}
	D = \tfrac{1}{3} N C = \tfrac{1}{12}N(N^2-1) \fineq{}
	\label{eq:rge:invariants}
\end{align}
for an SU(2) $N$-plet.
For the calculation of the RGEs, we will assume that the scalar potential is of the form given in \eqref{eq:twoScalars:potential}, \ie it is not only classically scale-invariant but additionally enjoys a global $O(4)\times O(N_\chi) \times O(N_\xi)$ symmetry.
The one-loop scalar-sector beta functions then turn out to be
\begin{align}
	\begin{split}
	\beta_{\lambda_\phi}^{(1)} ={}& 24\lambda_\phi^2 + 2 N_\chi \kappa_{\phi \chi}^2  + 2 N_\xi \kappa_{\phi \xi}^2 \fineq{,} \\
	\beta_{\lambda_\chi}^{(1)} ={}& 8(N_\chi+8) \lambda_\chi^2 + 2\kappa_{\phi\chi}^2 + 2 N_\xi \kappa_{\chi \xi}^2 \fineq{,} \\
	\beta_{\kappa_{\phi\chi}}^{(1)} ={}& 8\kappa_{\phi\chi} \Bigl[ \tfrac{3}{2}\lambda_\phi + (N_\chi+2)\lambda_\chi +\kappa_{\phi\chi} \Bigr] \\
	& + 4N_\xi \kappa_{\phi\xi} \kappa_{\chi\xi} \fineq{,} \\
	\beta_{\kappa_{\chi\xi}}^{(1)} ={}& 8\kappa_{\chi\xi} \Bigl[ (N_\chi+2)\lambda_\chi + (N_\xi+2)\lambda_\xi + 2\kappa_{\chi\xi} \Bigr] \\
	& + 4\kappa_{\phi\chi}\kappa_{\phi\xi} \fineq{.}
	\end{split}
	\label{eq:rge:ONscalarRGEs}
\end{align}
The beta functions of $\lambda_\xi$ and $\kappa_{\phi\xi}$ can be obtained from the ones of $\lambda_\chi$ and $\kappa_{\phi\chi}$ by exchanging $\chi \leftrightarrow\xi$ as well as identifying $\kappa_{\chi\xi}\equiv\kappa_{\xi\chi}$.

Taking into account the scalars' interactions with the electroweak gauge bosons, the above RGEs obtain the following additional contributions
\begin{align}
	\begin{split}
	\Delta\beta^{(1)}_{\lambda_\phi} ={}& -3\lambda_\phi (g_1^2 + 3g_2^2) + \tfrac{3}{8}(g_1^4 + 3 g_2^4 + 2 g_1^2 g_2^2) \fineq{,} \\
	\Delta\beta^{(1)}_{\lambda_\chi} ={}& -12 C_\chi \lambda_\chi g_2^2 + \tfrac{3}{32} \Bigl[  \mathcal{T}(N_\chi,N_\chi)  + 8 \delta_{N_\chi,3} \Bigr] g_2^4 \fineq{,} \\
	\Delta\beta^{(1)}_{\kappa_{\phi \chi}} ={}& -\tfrac{3}{2}\kappa_{\phi \chi} \Bigl[ g_1^2 + (4 C_\chi + 3) g_2^2 \Bigr] + \tfrac{3}{2} C_\chi g_2^4 \fineq{,} \\
	\Delta\beta^{(1)}_{\kappa_{\chi \xi}} ={}& -6 (C_\chi+C_\xi) \kappa_{\chi\xi} g_2^2 + \tfrac{3}{16} \mathcal{T}(N_\chi,N_\xi) g_2^4 \fineq{.}
	\end{split}
	\label{eq:rge:gaugeContrib}
\end{align}
where $C_\chi$ and $C_\xi$ denote the Casimir invariants of the scalar multiplets $\chi$ and $\xi$, respectively (\cf the definition in \eqref{eq:rge:invariants}). Furthermore, we use the abbreviation
\begin{align*}
	\mathcal{T}(N_\chi,N_\xi) = (N_\chi-1) (N_\xi-1) \Bigl[ N_\chi N_\xi - (N_\chi + N_\xi) + 3 \Bigr] \fineq{.}
\end{align*}
The only other sizable SM coupling is the top-quark Yukawa coupling $y$, which enters the scalar RGEs in the following way
\begin{align}
	\begin{split}
	\Delta\beta^{(1)}_{\lambda_\phi} = 6 (2 \lambda_\phi - y^2)y^2 \sepeq{,}
	\Delta\beta^{(1)}_{\kappa_{\phi\chi}} = 6 \kappa_{\phi\chi} y^2 \fineq{.}
	\end{split}
	\label{eq:rge:yukawaContrib}
\end{align}
Besides the Higgs beta function, the only SM RGE which changes in the presence of $\chi$ and $\xi$ is that of $g_2$, namely
\begin{align}
	\beta_{g_2}^{(1)} & = \left[ \tfrac{1}{6}(D_\chi + D_\xi) - \tfrac{19}{6} \right] g_2^3 \fineq{}
	\label{eq:rge:g2}
\end{align}
with the Dynkin indices $D_\chi$ and $D_\xi$ given in \eqref{eq:rge:invariants}.

\vspace{1em}
\subsection{SM + one complex scalar representation} \label{app:rge:SM+cplx}
\noindent
As in Section \ref{sec:oneScalar}, let $\chi$ be a complex scalar representation with $\chi\sim(1,\,N,\,Y)$ and let the scalar interactions be described by the potential in \eqref{eq:oneScalar:general_potential}. Note that we present the RGEs including $\kappa_3$, since we used them in our analysis in Section~\ref{sec:oneScalar:complex-woVEV}. The scalar sector beta functions are then calculated to be
\begin{align*}
	\beta_{\lambda_1}^{(1)} ={}& 24\lambda_1^2 + N \kappa_1^2
	+ \tfrac{1}{4} D \kappa_2^2 + 2D \kappa_3^2 \fineq{,} \\
	\beta_{\lambda_2}^{(1)} ={}& 4(N+4) \lambda_2^2 + 8 C \lambda_2\lambda_3 \\
	& + \tfrac{1}{2} (N-1)^2 \Bigl[ N (10-N) - 13 \Bigr] \lambda_3^2 \\
	& + 2\kappa_1^2 + \Bigl( \delta_{N,2} + \tfrac{9}{2} \delta_{N,4} \Bigr) \kappa_3^2 \fineq{,} \\
	\beta_{\lambda_3}^{(1)} ={}& \tfrac{1}{3} (N-2) \Bigl[ N(N+20)-33 \Bigr] \lambda_3^2 \\
	& + 24 \lambda_2 \lambda_3 + \tfrac{1}{2} \kappa_2^2 - 2\delta_{N,4} \kappa_3^2 \fineq{,} \\
	\beta_{\kappa_1}^{(1)} ={}& 4 \kappa_1 \Bigl[ 3\lambda_1 +(N+1)\lambda_2 + C \lambda_3 + \kappa_1 \Bigr] \\
	& + C \kappa_2^2 + 8 C \kappa_3^2 \fineq{,} \\
	\beta_{\kappa_2}^{(1)} ={}& 4\kappa_2 \Bigl[ \lambda_1 + \lambda_2 + (D+C -1) \lambda_3 + 2\kappa_1 \Bigr] - 16\kappa_3^2 \fineq{,} \\
	\beta_{\kappa_3}^{(1)} ={}& 4\kappa_3 \Bigl[ \lambda_1 + \lambda_2 - (C-1)\lambda_3 + 8\kappa_1 - 4\kappa_2 \Bigr] \fineq{,}
\end{align*}
where $N$, $C$ and $D$ refer to dimension and invariants of the representation under which $\chi$ transforms. The scalar sector is coupled both to SM fermions and to the electroweak gauge bosons. The corresponding contributions to the scalar RGEs are given by
\begin{align*}
	\Delta\beta_{\lambda_1}^{(1)} = 6 (2\lambda_1 -  y^2) y^2 \sepeq{,}
	\Delta\beta_{\kappa_i}^{(1)} = 6 \kappa_i y^2 \fineq{,}
\end{align*}
and
\begin{align*}
	\Delta\beta_{\lambda_1}^{(1)} ={}& -3 \lambda_1 \bigl( g_1^2 + 3 g_2^2 \bigr) + \tfrac{3}{8}g_1^4 + \tfrac{9}{8}g_2^4 + \tfrac{3}{4} g_1^2 g_2^2 \fineq{,} \\
	\Delta\beta_{\lambda_{2}}^{(1)} ={}& -3\lambda_{2} \Bigl[ 4 Y^2 g_1^2 + (N^2-1) g_2^2 \Bigr]  + 6Y^4 g_1^4 \\
	& + \tfrac{3}{8}(N-1)^2 \Bigl[N(10-N)-13 \Bigr] g_2^4 \fineq{,} \\
	\Delta\beta_{\lambda_{3}}^{(1)} ={}& -3\lambda_{3} \Bigl[ 4 Y^2 g_1^2 + (N^2-1) g_2^2 \Bigr] \\
	& + 12 Y^2 g_1^2 g_2^2 + 3\Bigl[ (N-3)^2-1 \Bigr] g_2^4\fineq{,} \\
	\Delta\beta_{\kappa_1}^{(1)} ={}&  -\tfrac{3}{2}\kappa_1 \Bigl[(4Y^2+1) g_1^2 +(N^2+2) g_2^2 \Bigr] \\
	&  + 12Y^4 g_1^4 + 3C g_2^4 \fineq{,} \\
	\Delta\beta_{\kappa_2}^{(1)} ={}&  -\tfrac{3}{2}\kappa_2 \Bigl[(4Y^2+1) g_1^2 + (N^2+2) g_2^2 \Bigr] \\
	& + 24 Y^2 g_1^2 g_2^2 \fineq{,} \\
	\Delta\beta_{\kappa_3}^{(1)} ={}&  -\tfrac{3}{2}\kappa_3 \Bigl[ (4Y^2+1) g_1^2 + (N^2+2) g_2^2 \Bigr] \fineq{.}
\end{align*}
The modified one-loop gauge RGEs are
\begin{align}
	\beta_{g_1}^{(1)} ={}& \Bigl( \tfrac{41}{6} + \tfrac{1}{3}N Y^2 \Bigr) g_1^3 \sepeq{,}
	\beta_{g_2}^{(1)} = \Bigl[ \tfrac{1}{3}D - \tfrac{19}{6} \Bigr] g_2^3 \fineq{.}
	\label{eq:rge:cmplxGauge}
\end{align}

\bibliographystyle{bibstyle}
\bibliography{RefMPIK-paper.bib}

\end{document}